%% file: main.tex
\documentclass[conference,10pt]{IEEEtran}
\IEEEoverridecommandlockouts

\usepackage{cite}
\usepackage{amsmath,amssymb,amsfonts}
\usepackage[skip=.4\baselineskip]{caption}
\usepackage[skip=.4\baselineskip]{subcaption}
\usepackage{graphicx}
\usepackage{hyperref}
\usepackage{textcomp}
\usepackage{xcolor}
\usepackage{subcaption} 
\usepackage{algorithm}
\usepackage{booktabs}
\usepackage[noend]{algpseudocode}
\usepackage{multirow}
\usepackage{minted} 
\usepackage{makecell}
\usepackage{soul}
\usepackage{xspace}
\usepackage{stfloats}
\usepackage{comment}
\usepackage{microtype}
\usepackage[subtle]{savetrees}
\usepackage[mode=text,group-four-digits=true,group-separator={,},exponent-product=\cdot,binary-units=true]{siunitx}
\DeclareSIUnit\flop{FLOP}
\newcommand{\mudock}{{muDock}\xspace}
\def\BibTeX{{\rm B\kern-.05em{\sc i\kern-.025em b}\kern-.08em
    T\kern-.1667em\lower.7ex\hbox{E}\kern-.125emX}}
\begin{document}

\newpage
\onecolumn                
\thispagestyle{empty} 
    \vspace*{2cm}
    {\LARGE \textbf{IEEE Copyright Notice}}\\[2cm]
    © 2025 IEEE. Personal use of this material is permitted.  Permission from IEEE must be obtained for all other uses, in any current or future media, including reprinting/republishing this material for advertising or promotional purposes, creating new collective works, for resale or redistribution to servers or lists, or reuse of any copyrighted component of this work in other works. \\[1cm]
    This work has been accepted for publication in CLUSTER 2025. The final published version will be available via IEEE Xplore
\clearpage
\twocolumn              

\setcounter{page}{1}

\title{Towards High-Performance and Portable Molecular Docking on CPUs through Vectorization}

\author{
\IEEEauthorblockN{Gianmarco Accordi\IEEEauthorrefmark{1}, 
Jens Domke\IEEEauthorrefmark{2}, 
Theresa Pollinger\IEEEauthorrefmark{2},
Davide Gadioli\IEEEauthorrefmark{1},
Gianluca Palermo\IEEEauthorrefmark{1}}

\IEEEauthorblockA{\IEEEauthorrefmark{1}Dipartimento di Elettronica, Informazione e Bioingegneria, Politecnico di Milano, Milan, Italy\\
\{gianmarco.accordi, davide.gadioli, gianluca.palermo\}@polimi.it}

\IEEEauthorblockA{\IEEEauthorrefmark{2}RIKEN Center for Computational Science, Kobe, Japan\\
\{jens.domke, theresa.pollinger\}@riken.jp}
}

\maketitle

\begin{abstract}
	Recent trends in the HPC field have introduced new CPU architectures with improved vectorization capabilities that require optimization to achieve peak performance and thus pose challenges for performance portability.
	The deployment of high-performing scientific applications for CPUs requires adapting the codebase and optimizing for performance.
        Evaluating these applications provides insights into the complex interactions between code, compilers, and hardware.
	We evaluate compiler auto-vectorization and explicit vectorization to achieve performance portability across modern CPUs with long vectors.
	We select a molecular docking application as a case study, as it represents computational patterns commonly found across HPC workloads.
	We report insights into the technical challenges, architectural trends, and optimization strategies relevant to the future development of scientific applications for HPC.
	Our results show which code transformations enable portable auto-vectorization, reaching performance similar to explicit vectorization.
	Experimental data confirms that x86 CPUs typically achieve higher execution performance than ARM CPUs, primarily due to their wider vectorization units.
	However, ARM architectures demonstrate competitive energy consumption and cost-effectiveness.
\end{abstract}

\input{sections/introduction.tex}
\input{sections/background.tex}
\input{sections/related_work.tex}
\input{sections/compilers.tex}
\input{sections/application.tex}
\input{sections/architectures.tex}
\input{sections/results.tex}
\input{sections/discussions.tex}
\input{sections/conclusions.tex}

\section*{Data and Code Availability}
\mudock{}~\cite{mudock} is an open-source mini-app, and our scripts for generating benchmark results, producing roofline plots, and setting up the environment are available on Zenodo~\footnote{\url{}{https://doi.org/10.5281/zenodo.14905033}}.



\bibliographystyle{IEEEtran}
\bibliography{main}

\end{document}

%% file: sections/introduction.tex
\section{Introduction}
\label{sec:introduction}
High-performance computing (HPC) is pivotal in solving large-scale scientific simulations~\cite{bigrun,vssummit}.
In HPC systems, CPUs are ubiquitous either as the main computing unit or as the host coordinator of the accelerators~\cite{TOP500}.
Notably, several supercomputing centers design their infrastructure to rely on CPU partitions, including Fugaku~\cite{fugaku}, LUMI-C~\cite{lumiC}, and ECMWF's Sequana XH2000 system~\cite{ecmwf_supercomputer}. 
Despite the growing adoption of GPUs in HPC, CPU vendors have continued to enhance processor performance by introducing new architectural features.
These include multi-core designs, advanced vector processing units, and deeper cache hierarchies with increased memory bandwidth~\cite{modernCPU}.
However, CPU usage can become a burden when targeting different instruction set architectures (ISA) or CPU configurations, which requires adapting, testing, validating, and tuning the code.
Despite steady advances in CPU design, fully leveraging their computing capabilities remains a challenge for domain scientists.

\begin{figure}[t] 
	\centering
	\includegraphics[width=0.85\columnwidth, keepaspectratio]{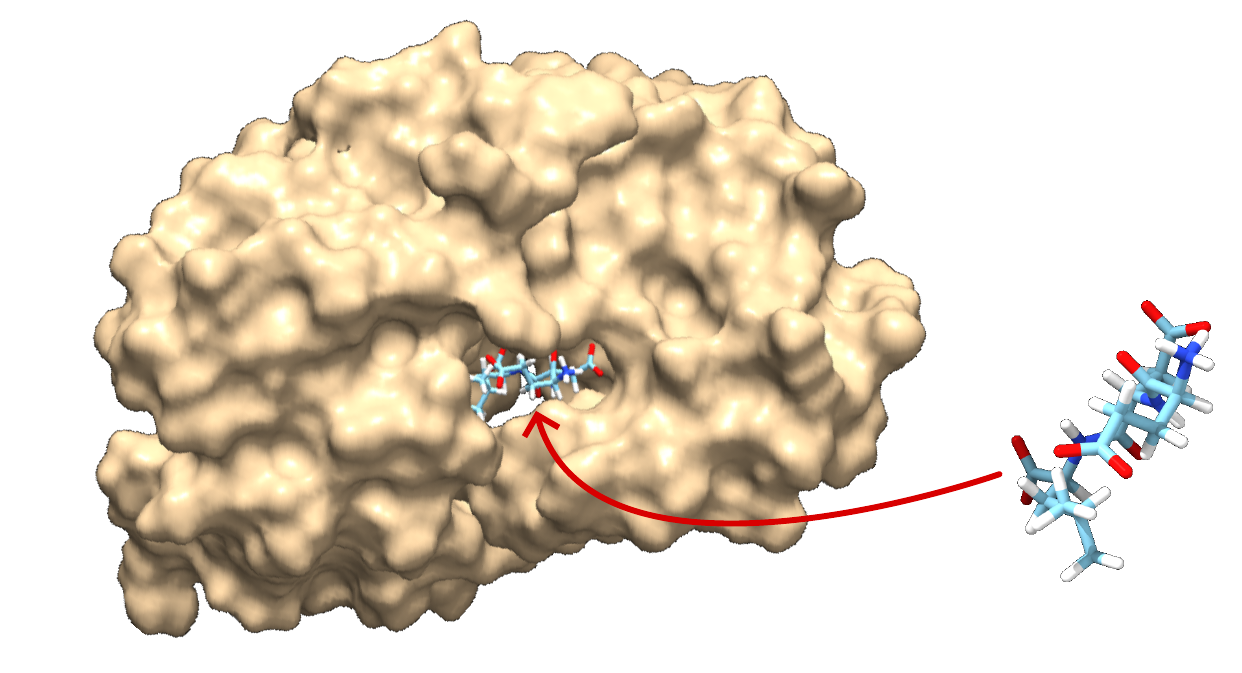}
	\caption{Visual representation of the docking problem. The ligand on the right gets docked in a protein-binding pocket.}
	\label{figure:complex}
\end{figure}

Biological applications consume a significant fraction of compute cycles at supercomputing centers~\cite{biousage}.
We focus on the molecular docking problems, which aim to estimate the preferred binding pose of a small molecule, named ligand, to a protein, as shown in Figure~\ref{figure:complex}.
Virtual screening applications need to solve this problem to deem a ligand a promising drug candidate.
For example, recent extreme-scale campaigns~\cite{bigrun,vssummit} used HPC resources to screen billions of ligands against the COVID-19 virus~\cite{Beccari2023}.
AutoDock~\cite{autodock} is one of such applications used in these campaigns.

To aid the scientific progress of computational biologists, we evaluate state-of-the-art auto-vectorization capabilities of commonly used compilers and compare them to explicit vectorization capabilities offered by portable frameworks such as Google Highway~\cite{googleHighway}.
In particular, we analyze two computational patterns in molecular docking: memory lookups into large constant data structures and loop vectorization~\cite{gromacs,openfoam}.
However, given the iterative approach of molecular docking algorithms, our findings might be transferable to other scientific applications.

We rely on a molecular docking mini-app called \mudock~\cite{mudock}, designed to assess the performance across architectures and portability frameworks, facilitating rapid prototyping.
Compared with other approaches reported in the literature, \mudock preserves the kernel complexity of well-known molecular docking frameworks (e.g, AutoDock), while offering high portability and extensibility.
Our work makes the following contributions:
\begin{itemize}
	\item We compare performance-portable approaches: compiler auto-vectorization and explicit vectorization for CPU; 
	\item We comprehensively evaluate modern CPUs for HPC using a realistic workload: a molecular docking code;
        \item We provide practical guidelines for code transformations of portable HPC docking applications;
	\item We discuss challenges in developing scientific codes with respect to compilers, vectorization, and memory hierarchy.
\end{itemize}

%% file: sections/background.tex
\section{Background}
\label{sec:background}
Virtual screening is an early stage of drug discovery, which is the process of finding suitable treatments against diseases.
Drug discovery is a lengthy and costly process~\cite{cost}, involving a considerable effort for pharmaceutical companies in terms of money and time.
The virtual screening process begins by identifying the main target, for example, a protein that leads to a disease, which is also called a receptor~\cite{moderndd}.
Given a target, the next stage of the process involves searching for potential molecules to interact with it.
Researchers then refine the most promising candidates before evaluating them in-vitro~\cite{moderndd}.
The initial search space in drug discovery may consist of billions of molecules.
This vast search space makes extensive in-vitro screening impractical.
Thus, any computational speed‑up that enables screening more molecules can increase the probability of drug discovery success~\cite{liu2025impact}.
One promising approach to address this challenge is using in-silico computational techniques~\cite{DDVS}.
Molecular docking is one of these approaches, where computational methods predict how small molecules (ligands) bind with a biological target.
The receptor's surface contains specific regions where molecule interactions are more likely.
These regions are known as binding sites.
Previous stages of the Drug Discovery pipeline, namely the Target Identification~\cite{moderndd}, identify the main structure of the protein.
For example, Alphafold~\cite{alphafold} is a tool to predict protein structure.

Molecular docking operates by placing a ligand within the binding site of a receptor, cf.~Figure~\ref{figure:complex}.
A scoring function then evaluates the interaction strength between the two~\cite{md}, estimating a binding affinity value. 
In this context, we define a pose as the spatial arrangement of a molecule upon docking, which includes rotations and translations within the receptor's binding site.
The set of admissible poses for each molecule-protein pair, also called \emph{complex}, is vast.
The scoring function assigns a numerical score to each pose, ranking them based on their predicted binding strength.
The ultimate goal of molecular docking is to efficiently explore the docking pose space and identify the most favorable poses associated with the best scores.

However, an exhaustive search is prohibitive due to the many possible poses and molecular configurations.
Molecular docking applications employ heuristics that guide a local search. 
These heuristics aim to balance computational efficiency and accuracy.
Therefore, molecular docking allows for identifying promising drug candidates without requiring impractical computational resources.
Accordingly, docking can provide acceptable results in a reasonable amount of time, assuming an adequate computing resource~\cite{md}.
Given the embarrassingly parallel nature of the problem, we can parallelize the workload on different execution units to speed up the process.
Literature reports examples of successful large-scale virtual screening campaigns with supercomputers~\cite{bigrun,minimdock}, in which one of the widely used molecular docking applications is AutoDock~\cite{autodock,autodock4,minimdock}.
%

%% file: sections/related_work.tex
\section{Related Work}
\label{sec:related_work}
Evaluating the performance of scientific applications on modern architectures is a long-standing topic of HPC.
Researchers have developed a wide range of benchmarking methodologies to assess both peak hardware capabilities or the performance of real-world workloads.
One of the most widely recognized benchmarks for HPC applications is LINPACK~\cite{linpack}, which includes a set of carefully tuned synthetic kernels designed to capture the peak \si{\flop/\second} performance of a platform.
Benchmark suites for real-case workloads, e.g., OpenBenchmarking~\cite{openbenchmarking}, benchpark~\cite{benchpark}, or SPEC CPU~\cite{speccpu}, exist, which offer a collection of open scientific applications for testing hardware performance.
Domain-specific suites, e.g., for genomics applications~\cite{GenArchBench}, also exist; however, we focus specifically on \mudock (and AutoDock) as representatives for the bioinformatics workloads on supercomputers.
Microbenchmarks offer more granular performance evaluations of specific CPU architecture features~\cite{memoryARM,micrograce}.
For instance, several studies have evaluated performance on the A64FX processor~\cite{10.1145/3636480.3637097,10.1145/3581576.3581621}.
Instead, we propose an evaluation of a real-case scientific workload, which better captures the complex interaction between code and compilers.

Recent research leveraging frameworks such as SYCL or MLIRSYCL~\cite{SYCLARM} has begun to evaluate performance portability on CPUs.
Portable tools like SYCL~\cite{kokkos} and Kokkos~\cite{kokkos} can target multi-core CPUs while providing portability.
However, vectorization is left to compilers' capabilities, while we go in-depth into them.

%

Several studies benchmark molecular docking applications to assess the performance of various algorithms and platforms~\cite{poenaru2021evaluation,minimdock}.
Some of these efforts have emphasized the importance of low-level optimizations, such as explicit vectorization techniques, to improve computational efficiency~\cite{ficarelliAsia}.
These works demonstrate that tailoring molecular docking implementations to leverage modern CPUs' capabilities can yield substantial speedups.
Studies often concentrate on specific computational kernels, e.g., Lennard-Jones potential evaluations, or they are limited to specific instruction set extensions, e.g., AVX~\cite{ljavx}, whereas we analyze diverse instruction sets.
GPU-based molecular docking frameworks have also received significant attention, with tools that explore the performance portability across different GPU platforms ~\cite{minimdock,poenaru2021performance}.
Nevertheless, there is less work systematically comparing CPU performance portability.

%% file: sections/compilers.tex
\section{CPU Portability and Tuning Approaches}
\label{sec:targetProblem}CPUs remain an important building block in HPC~\cite {TOP500}.
CPU vendors have increasingly adopted vector units and more sophisticated cache hierarchies, combined with higher core densities, to enhance performance.
HPC centers mainly deploy x86 CPUs~\cite{TOP500} from Intel and AMD, which support Advanced Vector Extensions (AVX) instructions of up to 512 bits~\cite{avx} to allow higher Single Instruction Multiple Data (SIMD) parallelism.
Furthermore, ARM is gaining market share, notably with Fugaku~\cite{fugaku} and NVIDIA's Grace chips~\cite{10.1145/3636480.3637097} tailored for HPC workloads.
Similarly to x86, ARM defines a Scalable Vector Extension (SVE) for SIMD computation~\cite{sve}, supporting vector lengths from 128 to \num{2048} bits.
Compiling for HPC systems requires careful selection of a toolchain consisting of various compilers and libraries.
The choice of compiler and compilation flags can significantly impact performance on a given architecture~\cite{ookami}.
Given the diverse landscape of available CPUs in the HPC field, as described before, scientific application deployment requires extensive optimization and tuning to achieve peak performance regarding vectorization units and memory hierarchy.
Developers typically follow two main approaches when adapting scientific applications for HPC CPUs: automatic or explicit vectorization.
Auto-vectorization~\cite{autovectorization} is a compiler optimization method that transforms scalar code into vectorized code.
Explicit vectorization requires developers to write architecture-specific code, which they can achieve using intrinsics or with performance portable libraries, like std::simd~\cite{stdsimd}, XSIMD~\cite{xsimd}, or Google Highway~\cite{googleHighway}.
We use Google Highway (HWY)~\cite{googleHighway} as a reference for explicit vectorization, against compiler auto-vectorization capabilities.
Additionally, HWY is a stable library and supports a wide range of CPU vectorized instruction sets, like AVX and SVE, as well as the mathematical functions required by \mudock computational kernels.
HWY emits optimized SIMD instructions based on the target architecture.
It supports SVE2 on ARM processors and AVX-512 on x86 processors.

We define performance portability as the ability of a single code base to attain efficient execution across diverse architectures without architecture-specific intrinsics.
Instead, we maintain a single code base and consider the compiler's ability to vectorize code via pragmas and optimization flags.
To assess the effectiveness of compiler auto-vectorization in achieving performance portability, we compare it against HWY.
Specifically, we analyze how well compilers can achieve vectorization on different architectures compared to HWY's explicitly optimized SIMD implementation, which we consider an architecture-specific baseline~\cite{pp}.

We consider auto-vectorization on the following compilers: LLVM, GCC, Fujitsu’s FCC, NVIDIA's NVC++, AMD's AOCC, and Intel's ICPX.
The target application is embarrassingly parallel since it can evaluate each input independently.
Therefore, we can compute more inputs in parallel rather than parallelize the computation of a single input.
This parallel computation avoids synchronization or communication issues and lets us focus on the vectorization problem.
Therefore, explicit parallelization, e.g., OpenMP~\cite{openmp} is unnecessary for our analysis.
Instead, we use a trivial work-stealing approach to parallelize computation inside a node.

Pragmas within the application's code influence compilers' auto-vectorization.
These pragmas provide hints to the compilers, thereby guiding them in their decision-making process regarding vectorization.
We use OpenMP pragmas to hint compilers' vectorization optimizations in this example.
\begin{samepage}
	\begin{minted}{cpp}
    #pragma omp simd
    for(...)
\end{minted}
\end{samepage}
OpenMP SIMD pragmas allow a portable way to guide the compiler toward aggressive vectorization optimization~\cite{openmp}.
While compiler-specific pragmas exist, e.g., \textit{\#pragma clang loop}, all compilers used in this study support OpenMP SIMD directives.
We observed that the OpenMP pragmas enable more aggressive optimization on specific architectures for GCC.
This is because the SIMD directives relax the compiler's heuristic cost model, often improving performance.
In all other cases, the OpenMP SIMD directives yield performance comparable to their respective compiler-specific pragmas.
Given the portability scope of this work, we adopt the OpenMP SIMD pragmas, to enable OpenMP optimizations without linking the OpenMP runtime.


%% file: sections/application.tex
\section{The molecular docking mini-app: \mudock}
\label{sec:application}

AutoDock~\cite{autodock4} is a widely used molecular docking application that serves as a foundation for domain experts to explore new features \cite{trott2010autodock}.
Due to the AutoDock relevance, its main computation kernels support GPUs~\cite{autodockGPU,autodockVINAGPU} and FPGAs~\cite{autodockFPGA}.
The primary goal of AutoDock is to improve and experiment with new approaches to solve the molecular docking problem.
Although AutoDock is modular and flexible, its full codebase complexity complicates isolated kernel analysis.
For this reason, we use \mudock~\cite{mudock}, an AutoDock-like mini-application designed for benchmarking and portability.
\mudock uses a genetic algorithm to dock a ligand inside the target protein binding site without a local search, similarly to the genetic algorithm used by AutoDock.
The scoring function defines an individual's fitness and is functionally equivalent to the one in AutoDock.
Given its embarrassingly parallel workload as AutoDock, extending \mudock{} to leverage multiple nodes using MPI is straightforward and out of the scope of this paper.

To evaluate an individual's fitness, i.e., to score a ligand, we need to apply the transformations encoded in the genotypes to the original ligand, i.e., dock the ligand.
\begin{algorithm}[t]
        \footnotesize
	\begin{algorithmic}[1]
		\State \textbf{Input:} Ligand, Population
		\State \textbf{Output:} Docked Ligands
		\For{individual \textbf{in} population} \label{alg:docking:for:individual}
		\State \texttt{translate\_ligand()} \label{alg:docking:translate}
		\State \texttt{rotate\_ligand()} \label{alg:docking:rotate}
		\For{bond \textbf{in} ligand} \label{alg:docking:for:fragment}
		\If{\texttt{rotatable(bond)}}
		\State \texttt{rotate\_fragments(bond)} \label{alg:docking:rotate:fragment}
		\EndIf
		\EndFor
		\EndFor
		\caption{Ligand-protein docking at each generation.}
		\label{alg:docking}
	\end{algorithmic}
\end{algorithm}
\begin{algorithm}[t]
        \footnotesize
	\begin{algorithmic}[1]
		\State \textbf{Input:} Docked Ligands, Protein
		\State \textbf{Output:} Docked Ligands Score
		\For{ligand \textbf{in} docked ligands} \label{alg:scoring:ligands}
		\For{atom \textbf{in} ligand} \Comment{Compute inter-energy} \label{alg:scoring:inter}
		\If{\texttt{nearby(atom, Protein)}}
		\State \text{score} += \text{\texttt{electrostatic(atom, Protein)}} \label{alg:scoring:inter:eletro}
		\State \text{score} += \text{\texttt{van\_der\_waals(atom, Protein)}} \label{alg:scoring:inter:vdw}
		\State \text{score} += \text{\texttt{hydrogen\_bond(atom, Protein)}}\label{alg:scoring:inter:hb}
		\State \text{score} += \text{\texttt{desolvation(atom, Protein)}} \label{alg:scoring:inter:desolv}
		\EndIf
		\EndFor
		\For{atom\_1 \textbf{in} ligand} \Comment{Compute intra-energy} \label{alg:scoring:intra}
		\For{atom\_2 \textbf{in} ligand}
		\If{\texttt{distance(atom\_1, atom\_2)}}
		\State \text{score} += \text{\texttt{electrostatic(source, dest)}} \label{alg:scoring:intra:eletro}
		\State \text{score} += \text{\texttt{van\_der\_waals(source, dest)}} \label{alg:scoring:intra:vdw}
		\State \text{score} += \text{\texttt{hydrogen\_bond(source, dest)}}\label{alg:scoring:intra:hb}
		\State \text{score} += \text{\texttt{desolvation(source, dest)}} \label{alg:scoring:intra:desolv}
		\EndIf
		\EndFor
		\EndFor
		\EndFor
	\end{algorithmic}
	\caption{Ligand-protein conformation scoring.~\cite{autodock}}
	\label{alg:scoring}
\end{algorithm}
Algorithm \ref{alg:docking} and Algorithm \ref{alg:scoring} emphasize the loop structure that drive vectorization..
In Algorithm~\ref{alg:docking}, we describe the docking algorithm applied at each generation.
It takes the ligand and the individuals encoded into the genotype as input.
For each individual, it performs a translation (Line~\ref{alg:docking:translate}) and rotation (Line~\ref{alg:docking:rotate}) according to the origin.
Then, it rotates one set of atoms connected to the rotamer, i.e., the rotamer fragment (Line~\ref{alg:docking:rotate:fragment}).
Algorithm~\ref{alg:docking} outputs docked ligands.

Algorithm~\ref{alg:scoring} describes the scoring algorithm that evaluates the interaction strength between the ligands and the protein (Line~\ref{alg:scoring:ligands}).
It takes the output of Algorithm~\ref{alg:docking} as input.
The docked ligand score is a combination of inter-energy and intra-energy contributions.
Each ligand's atom contributes to inter-energy computation (Line~\ref{alg:scoring:inter}), representing the interactions between ligand and protein atoms.
These inter-energy contributions include electrostatic, van der Waals, hydrogen bonding, and desolvation effects.
The contributions are computed based on the atom's positions and types.
The fixed structure of the receptor remains unchanged during the docking process, allowing us to precompute the inter-energy contributions through memoization and gridification~\cite{approx}: we precompute different 3D maps around the protein, as AutoGrid~\cite{autogrid} does.
AutoGrid (part of AutoDock) precomputes grid-based energy maps around the target receptor, using probe atoms of different types to calculate the potential interaction energies at various grid points.
For each point in space, the score contribution of a ligand's probe atom is calculated based on the atom's type and position.
AutoGrid constructs multiple layers of 3D maps, one for each possible input atom type, because the inter-energy contributions depend on the atom's position and type.
At runtime, a lookup in these grid maps is done based on the ligand's atoms' position and type (see Lines~\ref{alg:scoring:inter:eletro}--\ref{alg:scoring:inter:desolv}).
\mudock mimics this: it uses the input protein to precompute a set of grid maps for lookups in the inter-energy loop.

At Line~\ref{alg:scoring:intra}, we evaluate intra-energy contributions.
These contributions account for the internal ligand structure, meaning the forces between atoms of the ligand.
For each pair of ligand atoms (Line~\ref{alg:scoring:intra}), the algorithm evaluates their contributions to the ligand's score.
The intra-energy contributions include electrostatic, van der Waals, hydrogen bonding, and desolvation effects, similar to the inter-energy contributions.
However, since the ligand's conformation is not fixed (unlike the receptor), the intra-energy contributions cannot be efficiently precomputed.
Therefore, significant computation is required to calculate intra-energy contributions (Lines~\ref{alg:scoring:intra:eletro}--\ref{alg:scoring:intra:desolv}).
For example, using the Lennard-Jones formula, AutoDock~\cite{autodock} evaluates van der Waals and hydrogen bonding contributions.
This formula involves logarithmic and exponential computations for each pair of atoms in every ligand conformation at each docking generation.

The loop structure in \mudock facilitates benchmarking computational patterns that extend beyond molecular docking, making it applicable to a wider range of scientific workloads.
The inter-energy computation phase, represented by the loop at Line~\ref{alg:scoring:inter}, involves frequent lookups into constant memory.
This memory access pattern stresses the cache hierarchy and memory bandwidth, making it an ideal case for studying memory-bound performance bottlenecks.
Other applications in molecular simulations exhibit similar behavior, such as GROMACS, which uses tabulated interaction functions~\cite{gromacs} to run complex molecular dynamics simulations, or OpenFOAM~\cite{openfoam}, which utilizes precomputed lookup tables to speed up computations~\cite{openFOAMlt}.
In contrast, the intra-energy computation phase, shown by the loop at Line~\ref{alg:scoring:intra}, is characterized by computationally intensive mathematical operations.
This phase challenges the vectorization capabilities of both the CPU and the compiler, serving as an excellent test for compute-bound performance optimization.
Such a pattern is also typical in many scientific applications~\cite{lammps,wrf}.
This distinction between memory-bound and compute-bound operations makes \mudock a valuable tool for analyzing performance across diverse architectures.
To emit vector instructions, we need a minimum number of iterations to unroll, which depends on the type of data (e.g., float vs. double) and the width of the vectorization (e.g., 512-bit vs. 256-bit vs. 128-bit).
All the \mudock kernels meet this requirement.
In the case of microkernels that rely on recursive calls, the vectorization needs to hinge on the compiler to enable vectorization through code transformations, such as inlining.
For this reason, we consider them out of the paper scope.

%% file: sections/architectures.tex
\section{CPU Target Architectures}
\label{sec:architectures}
This section provides an overview of the CPU architectures we consider.
We categorize the architectures into two groups~\cite{GenArchBench}: ARM and x86.
We select x86 as the reference for traditional HPC computing because it is one of the most widely deployed CPU types in the Top500 list~\cite{TOP500}.
We evaluate AMD's Zen 4 (Genoa) and Intel's Golden Cove (Sapphire Rapids).
In addition, we include ARM CPUs from three vendors: Fujitsu, NVIDIA, and Amazon (AWS), due to their growing adoption~\cite{fugaku}.
Fujitsu's A64FX uses the ARMv8 architecture, and it was one of the first attempts at developing an ARM processor for exascale HPC~\cite{fugaku}.
We compare the A64FX with more recent variants: Graviton from Amazon and Grace from NVIDIA.
AWS and NVIDIA solutions employ the newer ARMv9 and Neoverse V2 cores.
We use AVX on x86 and SVE on ARM, cf.~Section~\ref{sec:background}.
Table~\ref{tab:cpuFeatures} summarizes key CPU specifications; Table~\ref{tab:cpuOOO} details out‑of‑order resources.
%
\begin{table*}[b]
	\centering
	\caption{Comparison of CPU Features.}
	\label{tab:cpuFeatures}
	\begin{tabular}{lcccccccccc}
		\toprule
		\multirow{2}{*}{\textbf{Vendor}} & \textbf{CPU}      & \multirow{2}{*}{\textbf{Architecture}} & \textbf{Max. Clock}  & \textbf{Max} & \textbf{Max} & \textbf{Vector}    & \textbf{Max}      & \textbf{Cost} & \multirow{2}{*}{\textbf{Year}} & \multirow{2}{*}{\textbf{Ref}}   \\
		                                 & \textbf{Codename} &                                        & \textbf{Speed} (\si{\giga\hertz}) &         \textbf{Core}*                       &                 \textbf{Thread}                 & \textbf{Extension} & \textbf{TDP} (\si{\watt})                 & (\$/NH)       &                                &                                 \\
		\midrule
		Intel                            & SPR   & Golden Cove                            & 4.8                  & 60                             & 120                              & AVX512             & 350                 & 3.82          & 2023                           & \cite{AWSPrice,sprSpec}         \\
		AMD                              & Genoa             & Zen 4                                  & 3.7                  & 96                             & 192                              & AVX512             & 400                 & 4.39          & 2022                           & \cite{AWSPrice,GenoaSpec}       \\
		AWS                              & Graviton 4        & Neoverse V2                            & 2.8                  & 96                             & 96                               & SVE2               & 130 & 3.40          & 2023                           & \cite{AWSPrice,neoverse}        \\
		Fujitsu                          & A64FX             & ARM Custom                             & 2.2                  & 48                             & 48                               & SVE2               & 150 & 0.64          & 2019                           & \cite{FugakuPrice,fujitsuA64FX} \\
		NVIDIA                           & Grace             & Neoverse V2                            & 3.4                  & 72                             & 72                               & SVE2               & 250                 & 11.17**        & 2022                           & \cite{alps,GraceGuide}          \\
		\bottomrule
		\multicolumn{8}{l}{*~Per socket~;~~~**~Compute node contains also a GPU}                                                                                                                                                                                                                                                                 \\
	\end{tabular}
\end{table*}
\begin{table}[b]
	\centering
	\caption{Comparison of CPUs out-of-order resources.}
	\label{tab:cpuOOO}
    \setlength{\tabcolsep}{3pt} 
        \resizebox{\columnwidth}{!}{
	\begin{tabular}{lccccccc}
    
		\toprule
		\textbf{Micro}        & \textbf{Instruction} & \textbf{Scalar}     & \textbf{Vector}      & \textbf{Vector} & \textbf{Vector}     & \multirow{2}{*}{\textbf{ROB}} & \multirow{2}{*}{\textbf{Ref}} \\
		\textbf{Architecture} & \textbf{Set}         & \textbf{Reg.}  & \textbf{Reg.} & \textbf{ALU}    & \textbf{Pipe.} &                               &                               \\
		\midrule
		Golden Cove           & x86                  & 288                 & 220                  & 512             & 2                   & 512                           & \cite{sprRegisters,spr}       \\
		Zen 4                 & x86                  & 224                 & 192                  & 256             & 1                   & 320                           & \cite{zen4}                   \\
		Neoverse V2           & ARMv9-A              & 213 & 188  & 128             & 4                   & 320                           & \cite{neoverse}               \\
		ARM Custom            & ARMv8.2-A            & 96                  & 128                  & 512             & 2                   & 128                           & \cite{fujitsuA64FX}           \\
		\bottomrule
		\multicolumn{7}{l}{\textit{NOTE: Number reported considering the maximum vector length}}                                                                                                                     \\
	\end{tabular}
    }
\end{table}

Genoa Zen 4 CPU features up to 12 Core Complex Dies (CCDs) with up to 8 cores in each CCD for 96 cores with two threads per core.
It supports AVX-512 instructions, even if vector units are 256 bits wide.
Thus, the Zen 4 process decomposes AVX-512 instructions into two 256-bit instructions.
The Zen 4 architecture can execute two separate 512-bit instructions in different pipeline stages.
The chip features a three-level cache hierarchy: \SI{64}{\kilo\byte} of L1 cache, \SI{1}{\mega\byte} of L2 cache, and 32/96 \si{\mega\byte} of L3 cache per CCD based on the specific chip.
Our Sapphire Rapids (SPR) CPU with Intel's Golden Cove cores supports up to 60 cores per socket, with two threads per core.
It has 512-bit vector registers and two 512-bit vector units and supports AVX-512 instructions.
As for the cache, it has an L1 cache of \SI{32}{\kilo\byte} for instructions and \SI{48}{\kilo\byte} for data, an L2 cache of \SI{2}{\mega\byte}, and a shared L3 cache of \SI{112.5}{\mega\byte}.
For comparison, Fujitsu's A64FX (based on ARMv8.2-A) operates at a lower maximum frequency of \SI{2.2}{\giga\hertz} and has an estimated TDP of \SI{150}{\watt}.
It features 128 512-bit wide vector registers and two vector pipelines.
A64FX consists of up to 48 cores (plus up to 4 OS assist cores in Fugaku), clustered into four core memory groups (CMG).
Each core has a \SI{32}{\kilo\byte} instruction cache and a \SI{32}{\kilo\byte} data cache, with \SI{8}{\mega\byte} of shared L2 cache per CMG.
We also evaluate the AWS Graviton 4 chip.
This processor has 96 Neoverse V2 cores running at \SI{2.7}{\giga\hertz} and supports SVE2 instructions with 128-bit wide registers.
It has four vector units, \SI{64}{\kilo\byte} L1 cache for instructions and data, \SI{2}{\mega\byte} L2 cache, and \SI{36}{\mega\byte} of shared L3 cache.
%
Lastly, we test NVIDIA's Grace chip~\cite{10.1145/3636480.3637097}, which features 72 Neoverse V2 cores per socket.
It supports SVE2 instructions, 128 bits wide, and four vector units.
Grace cores have a \SI{64}{\kilo\byte} L1 cache for instructions and data, a \SI{1}{\mega\byte} L2 cache per core, and a \SI{114}{\mega\byte} shared L3 cache.

As Table~\ref{tab:cpuFeatures} shows, the x86 CPUs generally sustain higher frequencies than ARM solutions but also have a higher TDP.
The number of cores and threads is similar across most architectures except for the A64FX processor, the oldest chip in our analysis.
On the other hand, the cost per node hour for the A64FX is significantly lower compared to more recent cloud solutions, with its energy consumption comparable to that of a Graviton chip.
We compute the node-hour cost for the A64FX based on Fugaku fees~\cite{FugakuPrice}, the Grace processor using Alps fees~\cite{alps}, and for all other CPUs, we use AWS's bare metal instance pricing~\cite{AWSPrice}.

Given that out-of-order execution resources are architecture-specific, we report them on a per-architecture basis.
Table~\ref{tab:cpuOOO} lists the available out-of-order resources for each architecture.
On average, x86 architectures provide more out-of-order resources than ARM processors, with the A64FX chip lagging behind other CPUs, particularly regarding the number of available registers and the reorder buffer (ROB) size.
The features most relevant to our analysis are each architecture's vector width and pipeline structure.
Only the Golden Cove and A64FX processors support native 512-bit vector instructions.
While Zen~4 also supports 512-bit vector instructions, it decomposes them into two 256-bit operations.
NVIDIA and AWS chips have smaller, 128-bit vector registers, though they compensate for this by offering more vector pipelines.
It is also worth noting that the A64FX processor includes the \textit{FEXPA} instruction, which allows for approximate computation of the exponential function, a crucial operation in \ muDock's inter-energy calculations.

%% file: sections/results.tex
\section{Experimental Setup}
\label{sec:setup}
We test the compilers and architectures described in Sections~\ref{sec:targetProblem} and~\ref{sec:architectures} using \mudock{}.
We conduct all experiments on a single testbed featuring all these CPUs, except that we use AWS cloud instances to access Graviton 4.

\paragraph{Hardware}
This section reports the features of the actual chip we used.
Frequency values reported here are the average value reported by LIKWID~\cite{likwid} during experiments across all cores.
Thus, some values may differ from those reported in Table~\ref{tab:cpuFeatures}.
The Xeon Platinum 8470 SPR CPUs operate at \SI{2.5}{\giga\hertz} in a dual-socket configuration with 52 cores per socket, totaling 208 threads.
Our AMD EPYC 9684X Genoa CPUs with Zen~4 cores are clocked at \SI{2.7}{\giga\hertz}, featuring 96 cores and 192 threads.
The A64FX CPUs are installed in air-cooled FX700 nodes, i.e., they operate at a maximum of \SI{2.0}{\giga\hertz} for all 48 cores.
We access the Grace Hopper (GH200) chip for NVIDIA's Grace CPU cores, which operate at \SI{2.5}{\giga\hertz} and feature 72 cores.
AWS's Graviton 4 cores are clocked at \SI{2.0}{\giga\hertz} and are available in a dual-socket configuration with 192 cores in total.

\paragraph{Software}
Intel, AMD, and NVIDIA nodes run Linux 5.14; Fujitsu FX700 uses 4.18, and Graviton 4 uses 6.1~\footnote{We lack administrative access to the testbed (shared among many active users), and hence unifying the kernel version was not within the paper's scope.
We performed multiple warm-up runs and observed low variance ($<1\%$) across repeats.}.

\begin{table*}[b]
	\centering
	\caption{Compiler versions and flags for x86 and ARM experiments.}
	\label{tab:compiler}
    \begin{tabular}{p{0.9cm}p{0.7cm}cc}
		\toprule
		\multirow{2}{*}{\textbf{Compiler}} & \multirow{2}{*}{\textbf{Version}} & \multicolumn{2}{c}{\textbf{Flags}}                            \\
		                                   &                                   & \textbf{x86}                       & \textbf{ARM}             \\
		\midrule
		LLVM                               & 19.1.0                            & \makecell{-fopenmp-simd                                       -ffast-math -fveclib=libmvec -march} & \makecell{-fopenmp-simd                           -ffast-math -fveclib=ArmPL -mcpu} \\
		\hline
		GCC                                & 15.0.0                            & \makecell{-fopenmp-simd -ffast-math -march}       & \makecell{-fopenmp-simd                           -ffast-math -mcpu}         \\
		\hline
		FCC                                & 4.11                              & N/A                                & \makecell{-Nclang -fopenmp-simd -ffast-math -mcpu}          \\
		\hline
		NVC++                              & 24.9                              & N/A                                & \makecell{-mp -Ofast -mcpu            }           \\
		\hline
		AOCC              & 5.0.0                             & \makecell{-fopenmp-simd -ffast-math -fveclib=AMDLIBM}                                & N/A                      \\
		\hline
		ICPX              & 2025.1.0                          & \makecell{-fopenmp-simd -ffp-model=fast}                                & N/A          \\
		\bottomrule
		\multicolumn{4}{l}{\textit{NOTE: march and mcpu are specific for the target architecture}}                                             \\
	\end{tabular}
\end{table*}

The two most widely used compilers we consider for auto-vectorization are LLVM (v19.1.0) and GCC (v15.0.0), built from upstream.
Additionally, we evaluate architecture-specific compilers from Fujitsu (FCC; v4.11), NVIDIA (NVC++; v24.9), AMD (AOCC; v5.0.0), and Intel (oneAPI compiler; v2025.1.0).
For FCC, we use \textit{clang mode} to compile our source code, which typically outperforms \textit{trad mode} for C++~\cite{fujitsuA64FX}.

We selected a small set of compiler flags relevant to compiler vectorization.
Table~\ref{tab:compiler} lists the compilation flags used in our experiments.
We use \textit{-ffast-math}, \textit{-O3}, and \mbox{\textit{-DNDEBUG}} to enable aggressive optimizations, such as auto-vectorization, relaxed floating-point arithmetic, and approximated mathematical functions.
To evaluate the impact of these aggressive optimizations, we compared docking scores from muDock with and without \textit{-ffast-math} on a subset of ligands, and the mean absolute difference in score was below $0.0002\%$.

The \textit{-march} and \textit{-mcpu} flags are architecture-specific, set to \textit{a64fx}, \textit{grace}, \textit{sapphirerapids}, \textit{znver4}, and \textit{neoverse\_v2}, respectively.
These flags optimize code generation and enable architecture-specific vectorized instruction sets.
Additionally, the \textit{-fveclib} flag enables the use of external performance libraries, facilitating vectorized mathematical operations on ARM architectures.
We use diagnostic flags, such as \textit{-fopt-info-vec-missed} for GCC and \textit{-Rpass-missed=loop-vectorize}, to get a report on the vectorization optimizations from compilers.
These flags provide reports on loops that failed to vectorize, highlighting the reasons for missed optimizations.

We fixed the result precision for performance measurement to type \textit{float} (32-bit). 
AutoDock also supports \textit{double}, but we expect similar performance trends with double, as vector instructions scale accordingly.

We collect results on \mudock with \num{100} individuals per population, over \num{1000} generations.
To parallelize the runs across cores, \mudock relies on \textit{pthreads}, allowing users to configure the number of cores used.
To optimize performance, \mudock sets thread affinity upon launching each thread, binding them to specific CPUs.
This approach helps mitigate cache invalidations and NUMA-related performance issues~\cite{numa}.

\paragraph{Vector-math libraries}
\mudock extensively relies on mathematical functions within its computational kernels.
For effective vectorization, compilers require explicit definitions of these functions.
Performance degrades significantly when the compiler fails to generate the appropriate vectorized instructions.
For instance, if the compiler lacks a vectorized exponential function, it will not vectorize the corresponding loops in \mudock.
Compilers leverage vectorization libraries to determine which functions they can optimize.
GCC relies on the vectorized libraries provided by the system's GLIBC.
Previous studies on ARM architectures~\cite{ookami} highlighted the absence of vectorized math functions in older GLIBC versions.
Recent GLIBC releases have improved support for vectorized functions on ARM, but vectorized GLIBC math functions are not always available on all systems.
As a result, some ARM system installations lack vectorized math libraries, such as \textit{libmvec.so}.
In this paper we have chosen to use the GLIBC version shipped with the OS for fairness toward explicit vectorization libraries (GH) or compilers with built-in math approximations (LLVM).
Indeed, HPC cluster users does not have the privileges to update GLIBC  at the system level. Moreover, compiling and correctly link a custom GLIBC is tecnhically challenging due to its integration with the OS and system libraries.
We use ARM Performance Libraries (ArmPL; v24.10) for LLVM, whereas for on x86 we rely on the system's GLIBC.
The Intel compilers and AMD's AOCC link their executable against their performance libraries, similar to ArmPL's approach.


Complex structures can hinder compiler auto-vectorization, e.g., deeply nested \textit{if} or \textit{for} statements.
Function calls within loops also impact vectorization, and hence we hint the compiler to maximize function inlining in \mudock, with \textit{inline} keyword and keeping kernels' helper functions in the same source file while enabling aggressive optimizations, with \textit{-O3}.
Additionally, we restructured certain mathematical computations, including reciprocal and exponential calculations, to leverage approximate computation optimizations~\cite{fujitsuA64FX},
However, generating the \textit{FEXPA} instruction was only possible using LLVM with ArmPL or the Fujitsu compiler.

\paragraph{Profiling}\label{sec:profiling}
We focus our experiments on the computational kernels of \mudock, specifically the docking and scoring steps described in Section~\ref{sec:application}.
%
In our experiments, we use LIKWID~\cite{likwid} to measure key metrics such as floating point operations (\si{\flop}), timing, memory bandwidth, energy consumption, and pipeline stalls.
To evaluate the peak performance of architectures, we use \textit{likwid-bench} across all available cores.
We use \textit{load} and \textit{peakflops} benchmarks tailored for each architecture.
For the energy consumption of Grace, we are using \textit{hwmon}, which utilizes the ACPI power meter~\cite{nvidiaPT}, and measures the CPU rail power for socket 0, excluding DRAM consumption.
AWS' Graviton 4 instances do not expose performance counters for power and memory; hence, we omit energy consumption and rooflines.

\paragraph{Dataset}
In our experiments, we used two input datasets derived from MEDIATE~\cite{mediate}, resulting from a large-scale virtual screening campaign~\cite{bigrun}.
The first dataset consists of around \num{2500} randomly selected molecules from the MEDIATE dataset and is used to evaluate performance across all available CPU cores.
We refer to this as the MEDIATE dataset.
The second dataset is designed for single-core execution and consists of the \textit{1a30} complex from PDBbind~\cite{pdbbind}.
We replicate the same molecule multiple times to consistently measure the computational kernel performance metrics across multiple runs.
Unless stated, metrics refer only to the inner kernels via LIKWID markers.

\section{Experimental Results}
\label{sec:results}
This section presents our experimental results\footnote{The plots use an NVCC label to indicate the NVC++ compiler.} to evaluate performance portability on modern CPUs.
We run experiments ten times, neglect the first three runs for each test, and average the metrics collected for the others.
The measured ratio between the standard deviation and the mean of experimental results is on average \SI{1}{\percent}.

\paragraph{Execution Time} The first experiment reports the execution time of \mudock{} in different scenarios.
The goal is to evaluate the impact of auto-vectorization on performance on diverse architectures against explicit vectorization.
We experiment with a single and a multi-core scenario.
We expect the CPU with wider vectorization units and out-of-order resources to be faster with respect to others.
We expect the multi-core scenario to follow a similar trend, accounting for the differences in core count across CPUs.
In both cases, the ideal case for auto-vectorized code is to perform similarly to HWY.

\begin{figure*}[t] 
	\centering
	\begin{subfigure}[t]{0.49\textwidth}
		\centering
		\includegraphics[width=\textwidth, keepaspectratio]{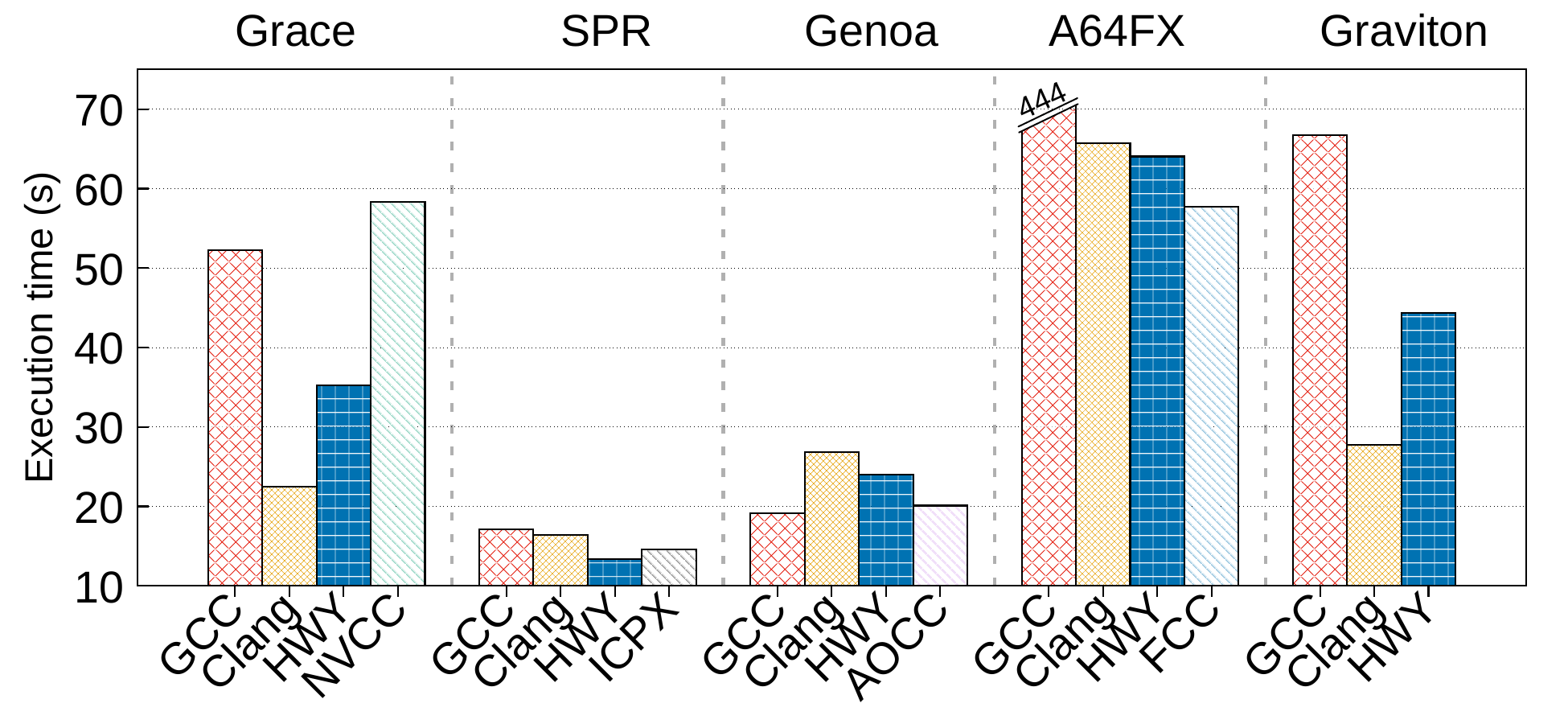}
		\caption{Single-core execution time with the reduced dataset.}
		\label{figure:time:single}
	\end{subfigure}
	\hfill
	\begin{subfigure}[t]{0.49\textwidth}
		\centering
		\includegraphics[width=\textwidth, keepaspectratio]{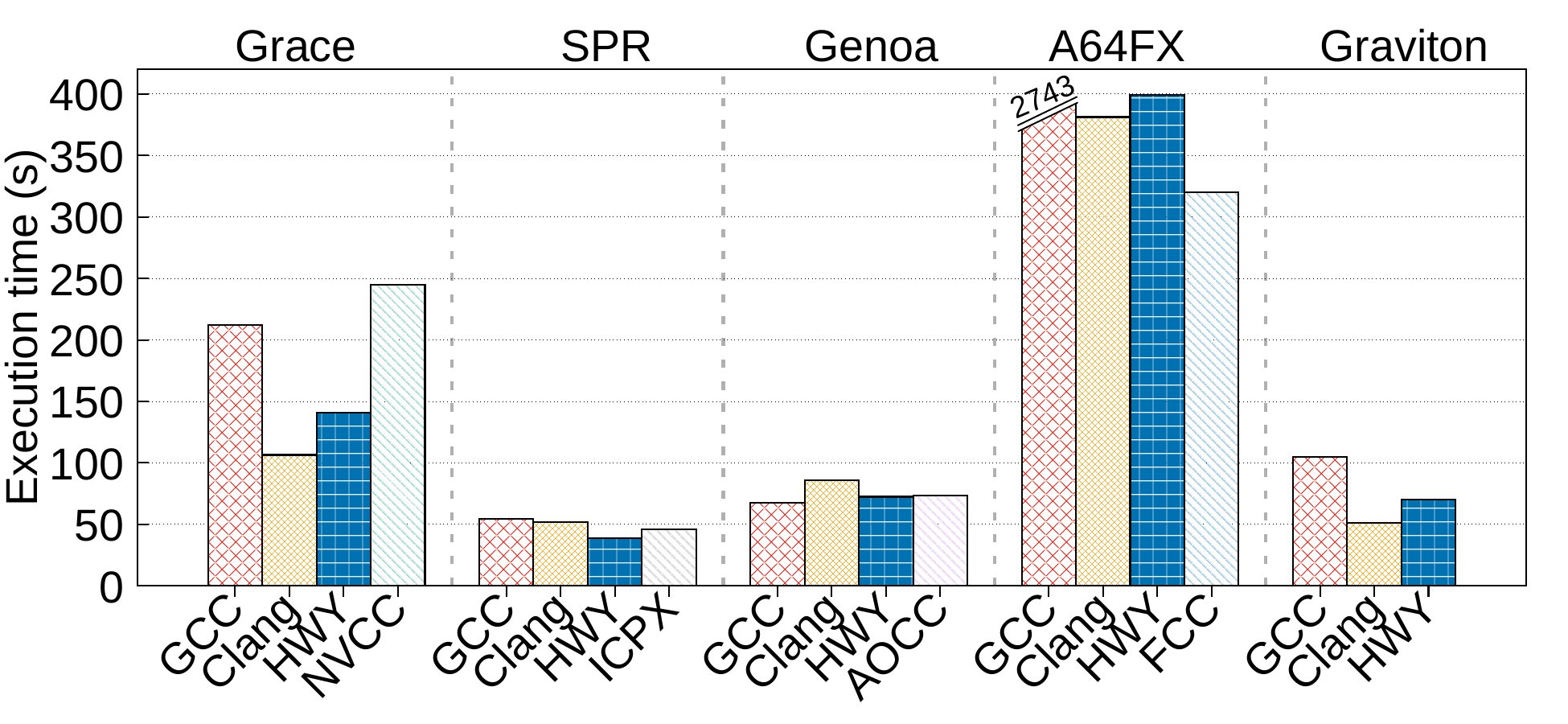}
		\caption{Full-core execution time with the MEDIATE dataset.}
		\label{figure:time:multi}
	\end{subfigure}
	\caption{Execution time of \mudock with a combination of datasets, compilers, and architectures.}
	\label{figure:time}
\end{figure*}
Figure~\ref{figure:time:single} reports single-core execution times on the reduced dataset of only \mudock kernels.
On the y-axis, it reports the execution times, while on the x-axis, it reports the differences in compilers, clustered by architecture.
As expected, Clang and HWY provide high performance across all architectures, with full support for all targets.
HWY performs better than Clang on architectures with a wider vectorization unit (512 bits), given that it can always vectorize to the maximum width.
On Neoverse architectures (128 bits), Clang performs better than HWY. 
Clang and HWY can use the maximum vector width, while Clang can deliver higher performance due to the usage of the ARM performance libraries.
As anticipated, GCC performance is slower on ARM architectures given the lack of a GLIBC vectorized version.
On Grace, NVC++ (NVCC) shares the same GCC issue, as it cannot vectorize loops, thus resulting in lower performance.
Interestingly, on the A64FX processor, we observed that the FCC version performed better than all others.
The FCC can emit specific A64FX approximated instructions, such as \textit{FEXPA}, and better leverage A64FX out-of-order resources and memory hierarchy.
Vendor-specific compilers, like ICPX and AOCC, link their executable against their performance libraries to link mathematical functions inside vectorized loops.
Despite that, we can see that they are not the fastest available solutions on their target platform.
GCC builds a faster executable on Genoa owing to fewer Last Level Cache (LLC) miss rates than other compilers, and a more aggressive usage of the cost model.
On SPR, HWY achieves the best performance, due to a limitation in the LLVM and GCC cost models.
The cost models restrict vectorization to 256-bit wide instructions to avoid frequency scaling issues observed on previous architectures~\cite{fscaling}.
In contrast, HWY explicitly emits 512-bit wide instructions, fully leveraging the available vector width on SPR.

Figure~\ref{figure:time:multi} reports the execution time only of \mudock kernels on all threads, using the MEDIATE dataset.
On the y-axis, it reports the execution times, while on the x-axis, it reports the differences in compilers, clustered by architecture.
As expected, the multi-core scenario shows a similar trend to the single-core one.
CPUs with a higher core count benefit from the parallel execution pattern offered by the molecular docking application.
Particularly, Graviton shows comparable performance to Genoa chips (96 vs 192 physical cores).
Meanwhile, CPUs with smaller core counts and out-of-order resources, like A64FX and Grace, show slower performance.

\begin{figure*}[t]
	\centering
	\includegraphics[width=1.4\columnwidth]{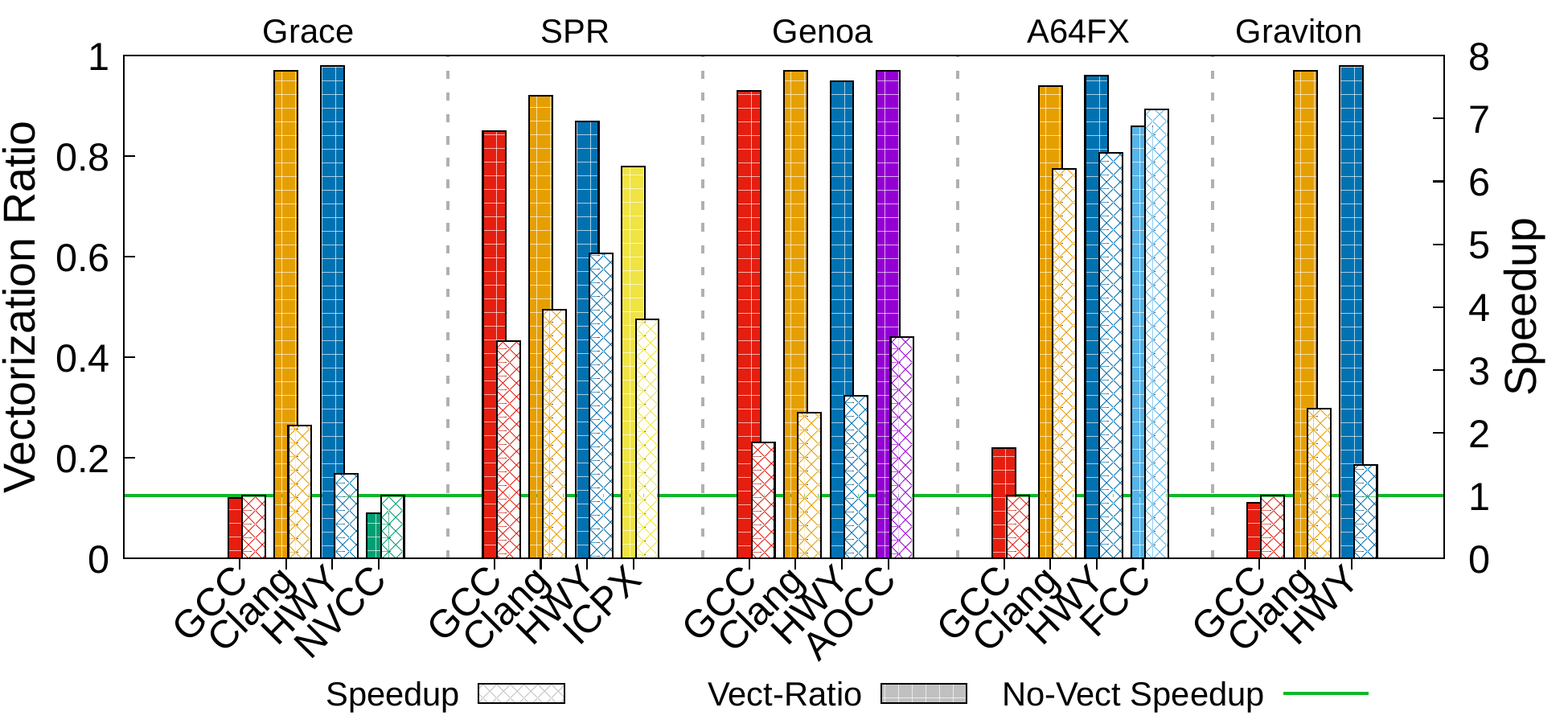}
	\caption{Correlation between speedup and vectorization ratio across diverse compilers and architectures.}
	\label{figure:speedupRatio}
\end{figure*}
To further investigate the differences in performance among compilers, we report in Figure~\ref{figure:speedupRatio} the impact of vectorization in terms of vectorization and speedup.
We define speedup as the ratio between the execution time with no vectorization and with vectorization on the same architectures using the same compiler (Clang for HWY).
We consider the vectorization as the ratio between scalar and vectorized instructions.
When measuring the vectorization ratio, we aggregated the 128-bit, 256-bit, and 512-bit instructions without distinguishing between widths.
We expect architecture with wider vectorization units to get a higher speedup and achieve a vectorization ratio close to \num{1} when compilers can vectorize.
We also expect the speedup to be comparable to the same architectures between vectorized code.
Figure~\ref{figure:speedupRatio} shows the performance of only \mudock's kernels, on a single-core execution of the reduced dataset.
On the left y-axis, it reports the vectorization ratio, while on the right y-axis, it reports the speedup ratio.
On the x-axis, it reports the differences in compilers, clustered by architecture.
On each entry on the x-axis, the bar on the left reports the vectorization ratio, while the bar on the right reports the achieved speedup.
The horizontal line in Figure~\ref{figure:speedupRatio} refers to the speedup axis, highlighting cases where the implementation over the non-vectorized code, like GCC on ARM architectures, achieves no speedup.
As expected, when compilers can auto-vectorize code, they get a vectorization ratio comparable to HWY's.
The only exceptions are GCC on ARM and NVCC on Grace, due to the missing vectorized GLIBC math libraries.
As expected, the speedup on architectures with support for 512-bit-wide instructions is higher, like SPR, Genoa, and A64FX.
Genoa shows the smaller speedup, given that it uses a 256-bit vector-wide unit instead of a 512-bit one like SPR and A64FX.
Genoa and SPR have a minor speedup than A64FX due to the usage of the \textit{SSE} instruction sets.
x86 processor uses the \textit{SSE} instructions set~\cite{sse}, which performs floating point arithmetic and consists of 128-bit-wide registers.
Since disabling \textit{SSE} instructions was impossible, the baseline for measuring the speedup on x86 uses vectorized instructions.

\begin{figure}[t]
	\centering
	\includegraphics[width=1.0\columnwidth]{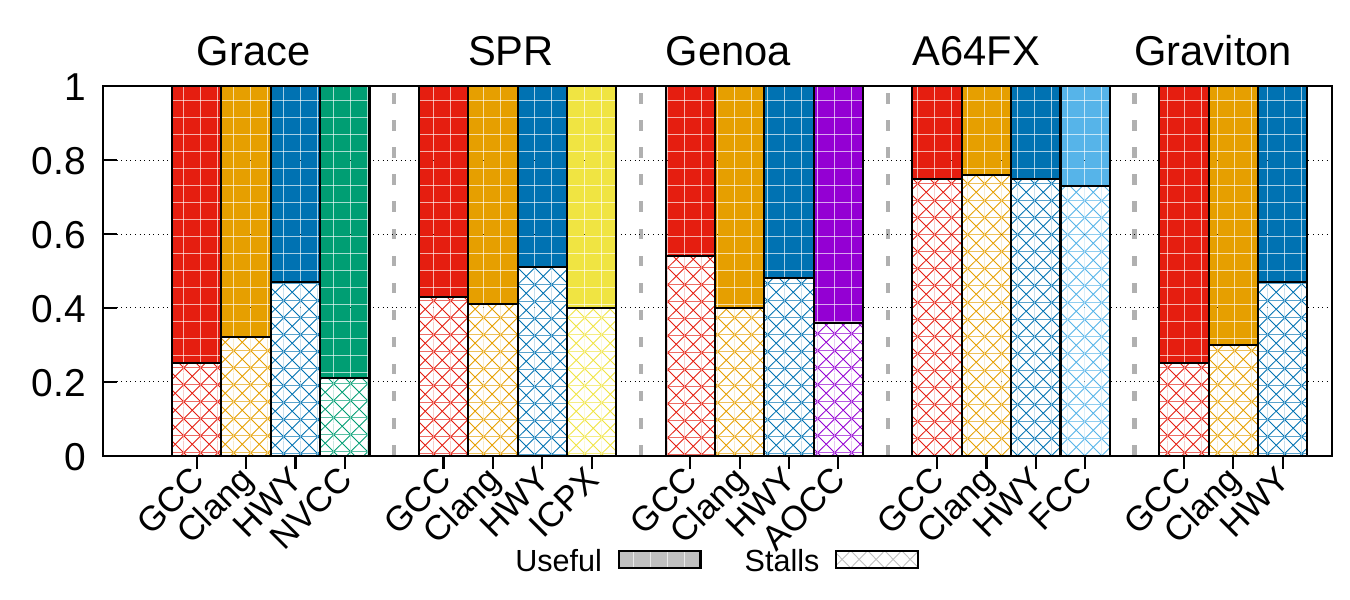}
	\caption{Ratio between pipeline stalls and useful work across diverse compilers and architectures.}
	\label{figure:stalls}
\end{figure}
Given the varying performance of the A64FX processor compared to other more recent ARM processors, we conducted a follow-up experiment to investigate the issue further.
This experiment aims to evaluate the impact of stalls on the execution pipeline.
Specifically, we focus on frontend and backend stalls, considering all other cycles useful work.
Therefore, we report only the total number of stalls, as our goal is to assess how the availability of out-of-order resources affects pipeline performance.
We expect to observe more stalls on the A64FX processor owing to its reduced availability of out-of-order resources.
Figure~\ref{figure:stalls} reports the experimental results only of \mudock's kernels, conducted on a single-core using the reduced dataset.
As shown in the figure, on average, \SI{70}{\percent} of the execution pipeline on the A64FX processor is occupied by stalls.
When we analyzed the A64FX two vector pipelines, we noticed how it uses the whole vector width available, but its busy rate is around \SI{20}{\percent}.
The number of stalls combined with the high miss-rate of the last level cache suggests that stalls are due to memory latencies in higher cache hierarchy levels.
The A64FX experiences a significantly higher number of stalls than other ARM architectures, such as Grace and Graviton, which have out-of-order resources comparable to x86 architectures.
The A64FX's limited out-of-order execution resources contribute to these stalls, particularly in intensive loops involving mathematical functions, which become bottlenecks.
A potential solution to mitigate this is loop fission, which reduces the resources required~\cite{loopfission}.
Other optimizations specific to the A64FX, such as those suggested for LLVM~\cite{a64fxSWP}, could address this issue, but we could not test them within this work's scope.
However, we do not consider the above solutions to be performance-portable.

\paragraph{Architectural Efficiency}
We evaluate the impact of vectorization on the architectural efficiency, which we define as the compiler's capabilities to use architectural resources efficiently.

\begin{figure} 
    \centering
    \begin{subfigure}[t]{0.31\textwidth}
        \centering
        \includegraphics[width=\textwidth, keepaspectratio]{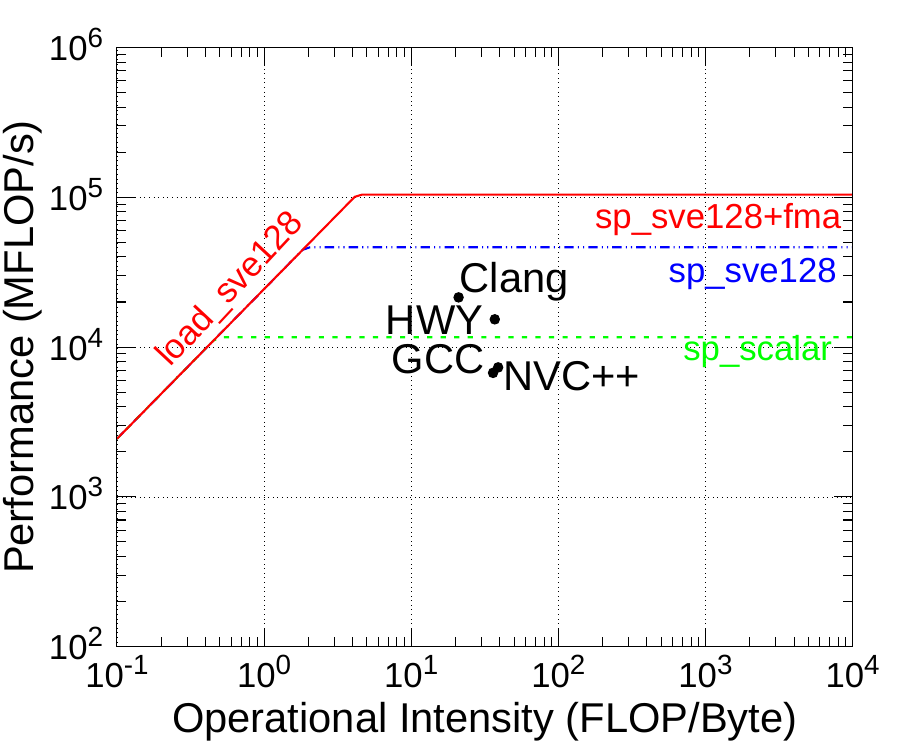}
        \caption{Rooflines for Grace}
        \label{figure:rooflineGrace}
    \end{subfigure}
    \hspace{0.01\textwidth}
    \begin{subfigure}[t]{0.31\textwidth}
        \centering
        \includegraphics[width=\textwidth, keepaspectratio]{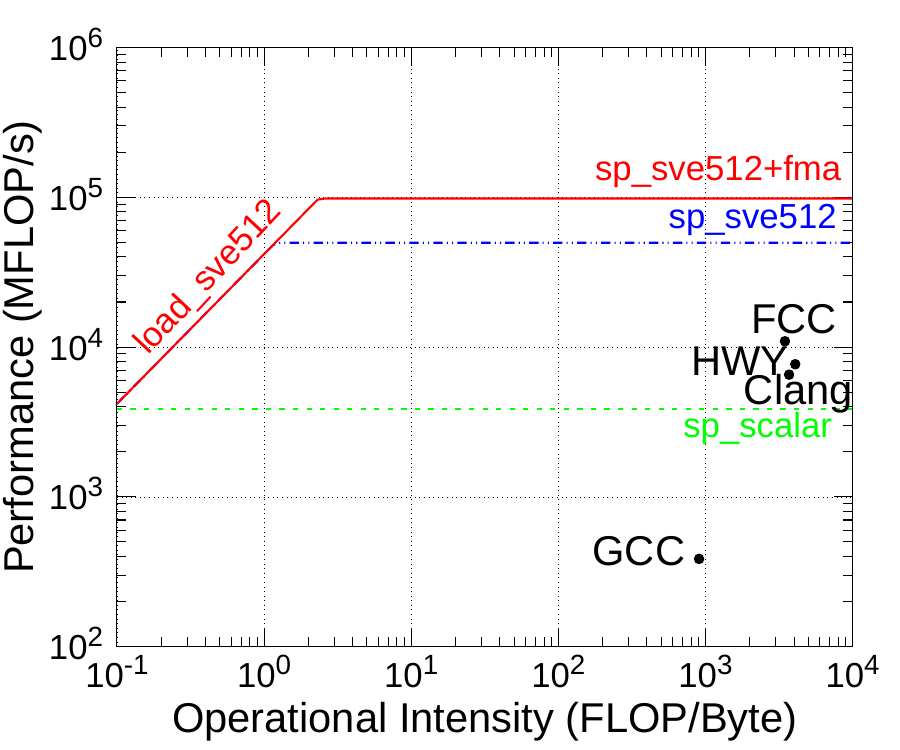}
        \caption{Rooflines for A64FX}
        \label{figure:rooflineA64FX}
    \end{subfigure}
    \hspace{0.01\textwidth}
    \begin{subfigure}[t]{0.31\textwidth}
        \centering
        \includegraphics[width=\textwidth, keepaspectratio]{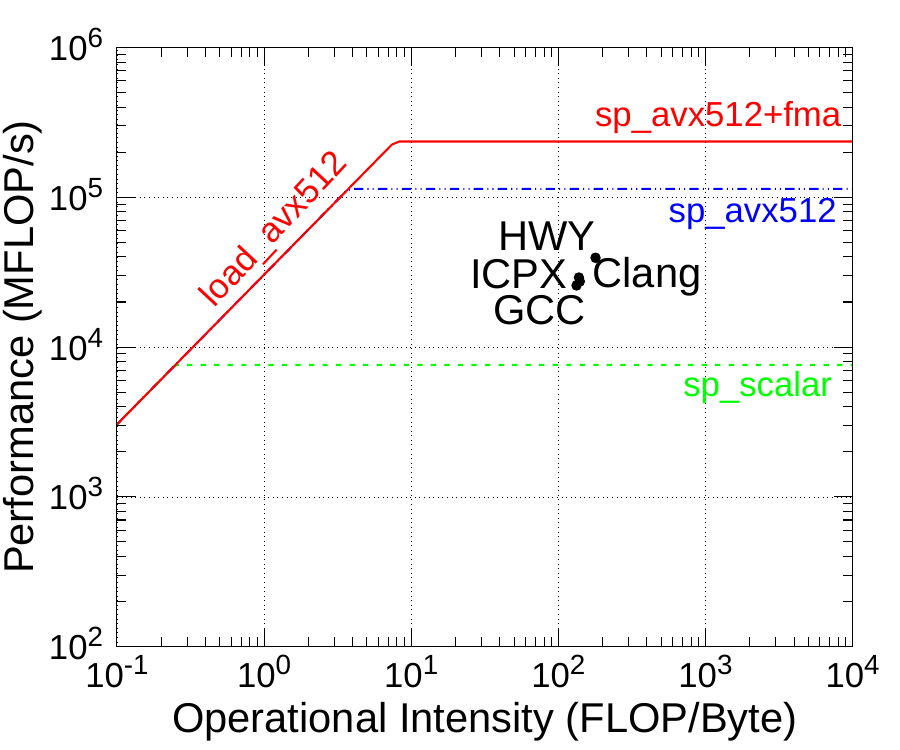}
        \caption{Rooflines for SPR}
        \label{figure:rooflineSPR}
    \end{subfigure}
    \hspace{0.01\textwidth}
    \begin{subfigure}[t]{0.31\textwidth}
        \centering
        \includegraphics[width=\textwidth, keepaspectratio]{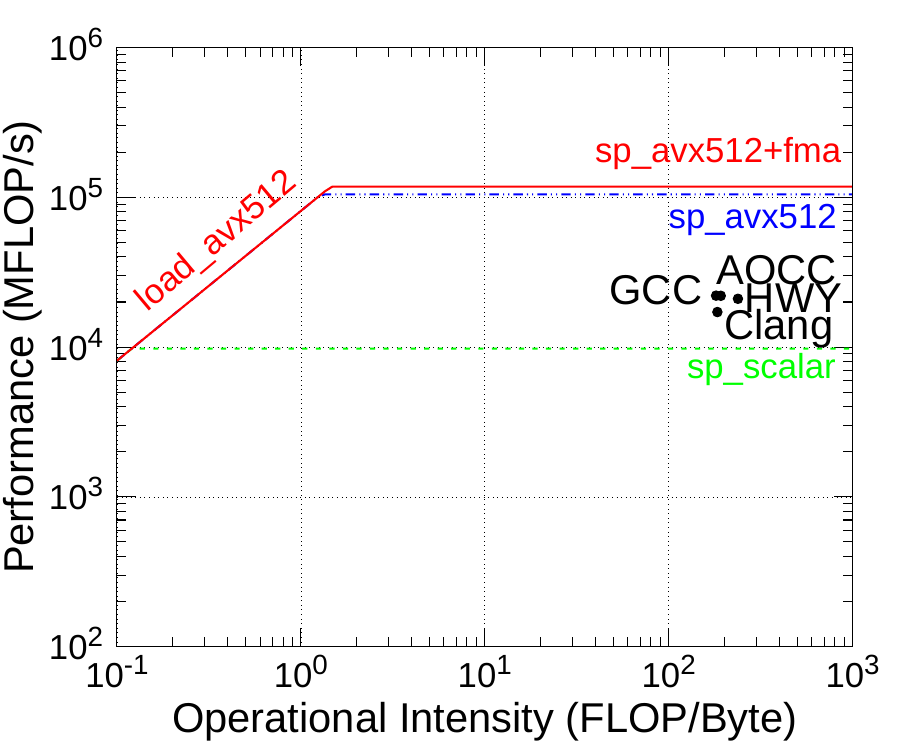}
        \caption{Rooflines for Genoa}
        \label{figure:rooflineGenoa}
    \end{subfigure}

    \caption{Kernel's rooflines on different architectures. All plots report data on a logarithmic scale.}
    \label{figure:rooflines}
\end{figure}

In Figure~\ref{figure:rooflines} we report \mudock's kernels' aggregated performance on diverse architectures, on a single-core run with the reduced dataset.
Each plot shows the compute performance on the y-axis in \si{\mega\flop/\second}, while the x-axis represents the arithmetic intensity in \si{\flop}/Byte.
Both axes use the logarithmic scale.
The peak memory bandwidth refers to the main memory bandwidth.
Graviton architecture is not reported, given that we could not access the hardware counter relative to memory bandwidth and last-level cache usage.
The experiment's goal is to evaluate if the application is memory or compute bound, while also visualizing how much \mudock is close to the peak performance.
Given the high arithmetic intensity described in Section~\ref{sec:application}, we expect the application to be compute-bound.
As expected, the roofline analysis confirms the high arithmetic intensity of \mudock, indicating that the application is compute-bound.
On average, the HWY and Clang implementations deliver strong performance across all architectures.
On x86 (Figures~\ref{figure:rooflineSPR} and~\ref{figure:rooflineGenoa}), all implementations perform well, as the vectorization permits reaching higher \si{\mega\flop/\second} values.
ARM shows a similar trend, except for GCC and NVCC compilers, given the aforementioned issue with GLIBC.
In particular, FCC on the A64FX improves the arithmetic intensity by emitting architecture-specific instructions, like \textit{FEXPA}.

\begin{table}[b]
	\centering
	\caption{Last Level Cache miss-rate across diverse architectures for Clang.}
	\label{tab:missrate}
	\begin{tabular}{l|cccc}
		\toprule
            \textbf{Mode} & \textbf{Grace} & \textbf{SPR} & \textbf{Genoa} & \textbf{A64FX}\\
		\midrule
            Single-core&\num{1.0e-4}&\num{2.03e-7}&\num{8.71e-5}&\num{6.92e-6}\\
            Multi-core&\num{3.38e-4}&\num{1.04e-5}&\num{2.08e-2}&\num{7.23e-4}\\
		\bottomrule
	\end{tabular}
\end{table}
\begin{table}[b]
	\centering
	\caption{Arithmetic intensity across diverse architectures for Clang.}
	\label{tab:ai}
	\begin{tabular}{l|cccc}
		\toprule
            \textbf{Mode} & \textbf{Grace} & \textbf{SPR} & \textbf{Genoa} & \textbf{A64FX}\\
		\midrule
            Single-core&\num{21}&\num{133}&\num{184}&\num{3700}\\
            Multi-core&\num{9313}&\num{12762}&\num{96}&\num{34}\\
		\bottomrule
	\end{tabular}
\end{table}
To investigate how compiler-architecture interactions impact memory hierarchy usage, we investigate the Arithmetic Intensity (AI) and LLC between single and multi-core cases.
For simplicity, we consider data coming from Clang, as other compilers show the same behavior.
Table~\ref{tab:ai} reports the AI of Clang implementation across diverse architectures, in a single and a multi-core scenario.
Table~\ref{tab:missrate} reports the LLC miss rate intensity of Clang implementation across diverse architectures, in a single and a multi-core scenario.
The Genoa architecture shows a lower AI than the Sapphire Rapids architecture on a multi-core execution with respect to the single-core ones.
The inter-energy computation accesses precomputed grid maps shared across all cores, which pressures the cache hierarchy.
Such a memory access pattern reduces the arithmetic intensity if the data is unavailable in the cache.
We expect performance to be slower on architectures that do not have a shared LLC or have higher latencies to main memory.
As expected, we notice a spike in the LLC miss rate between the single and the multi-core cases in Genoa.
On the Genoa CPU, the LLC cache is shared only by cores within the same CCD so that each core can reuse, at most, the precomputed values loaded by surrounding cores.
The single-core execution can fit data in the cache and prefetch it when needed, which is not true in the multi-core scenario.
Also, the miss-rate is higher than the other x86 architectures SPR, which shows a higher miss-rate in the multi-core scenario, with a high arithmetic intensity, given that the LLC is shared among all threads.
Grace shows a similar behavior to SPR, as its arithmetic intensity increases in the multi-core scenario, as Grace also has a shared LLC across cores.
On A64FX, each Core Memory Group uses its shared 8 MiB L2 as the last-level cache (there is no separate L3).
We measure the memory data transfer to compute the arithmetic intensity between the last-level cache and the main memory across architectures.
In a multiple-core scenario, we noticed a spike in L2 misses.
The A64XFX cannot accommodate all requests to the shared L2 cache from its group's cores, causing more L2 evictions and, thus, more HBM2 accesses.
Consequently, the arithmetic intensity drops~\cite{scagrace}.

\paragraph{Performance Portability}
\begin{figure*}[t] 
	\centering
	\includegraphics[width=1.3\columnwidth, keepaspectratio]{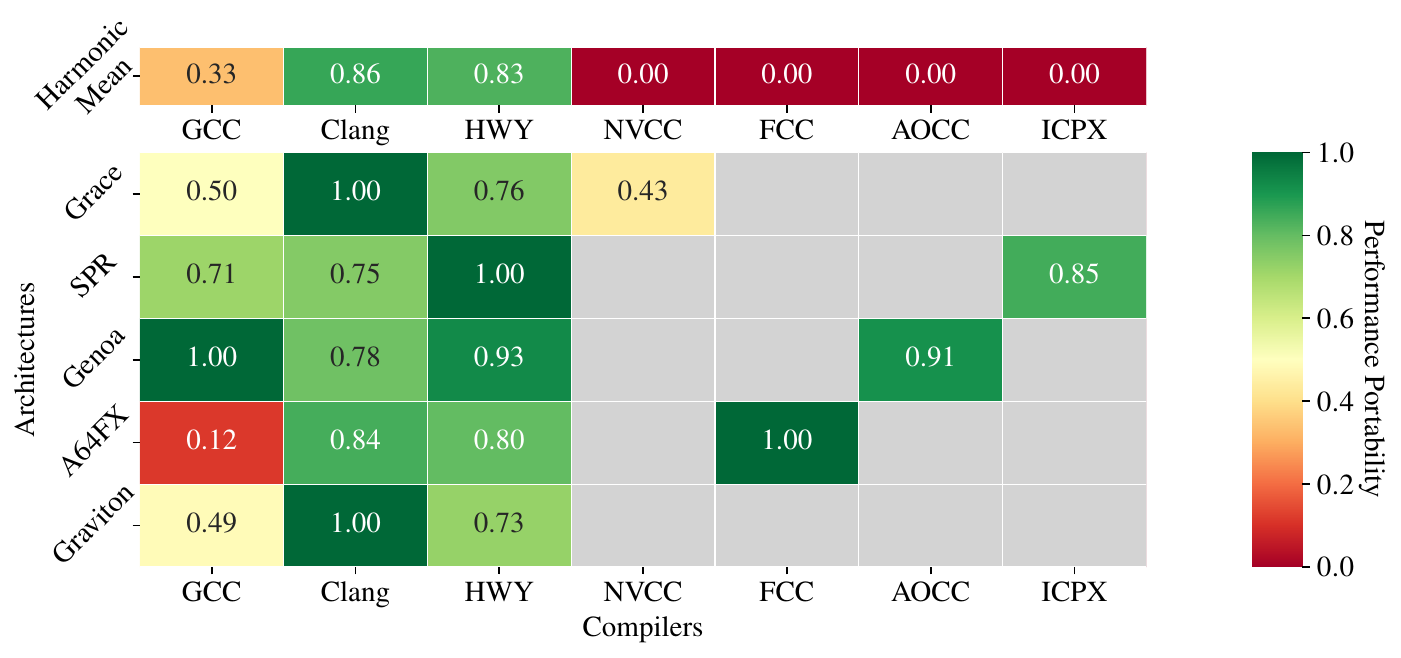}
	\caption{Compilers' application performance portability; gray indicates a lack of meaningful data for that compiler/architecture combination.}
	\label{figure:pp}
\end{figure*}
We compared how different solutions perform across architectures regarding performance portability to provide practical guidelines.
This experiment aims to visually highlight considerations on the interaction between the compiler and diverse architectures.
We define performance portability as the compiler's ability to achieve consistent peak performance across multiple architectures using a single codebase~\cite{pp}.
We treat compile-time flags (e.g., ffast-math, O3) and vendor libraries (e.g., ArmPL) as part of the build environment and outside performance portability considerations, as long as they do not involve code changes.
The idea is that building the toolchain from scratch and setting specific compiler presets is straightforward with modern build systems.
On the other hand, creating different code variants implies a significant degradation of maintainability.
Figure~\ref{figure:pp} shows the normalized execution time for the combination of compiler and architecture that we are considering.
We report the normalized execution time of a compiler implementation on an architecture against the best execution time on the same architecture (application efficiency).
On top, we report the performance portability as a harmonic mean~\cite{pprevised}.
GCC shows low-performance portability, especially on ARM architecture, given the vectorization issue with the GLIBC.
Because of the lack of vectorization, the GCC portability problem on ARM is more pronounced on the A64FX CPU.
These portability differences result from the discrepancy in vector width on other ARM architectures (512 vs 128 bits), which increases the gap in performance between a scalar and a vectorized code.
Clang provides performance portability comparable to HWY one: HWY performs better on x86, while Clang, with the ARM performance library, provides better performance portability on ARM architectures.
For example, A64FX HWY does not provide faster implementation, as FCC and Clang can better leverage architectural features and performance libraries.
As expected, the performance portability of FCC, NVC++, ICPX, and AOCC is $0$, which we consider platform-specific.

\paragraph{Computational Cost}
\begin{figure*}[t] 
	\centering
	\begin{subfigure}[t]{0.49\textwidth}
		\centering
		\includegraphics[width=1.0\columnwidth]{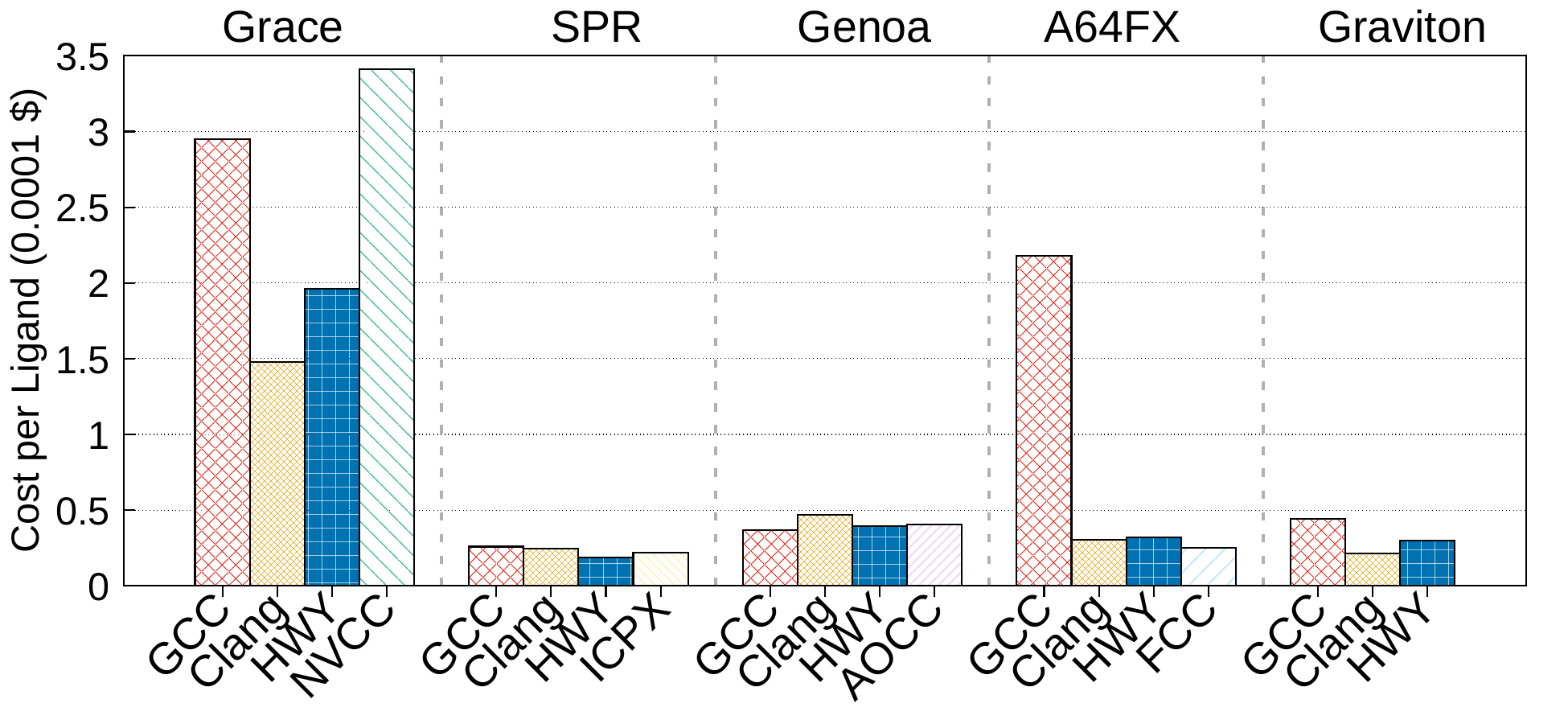}
		\caption{\mudock cost per ligand calculation (in USD).}
		\label{figure:cost}
	\end{subfigure}
	\hfill
	\begin{subfigure}[t]{0.49\textwidth}
		\centering
		\includegraphics[width=1.0\columnwidth]{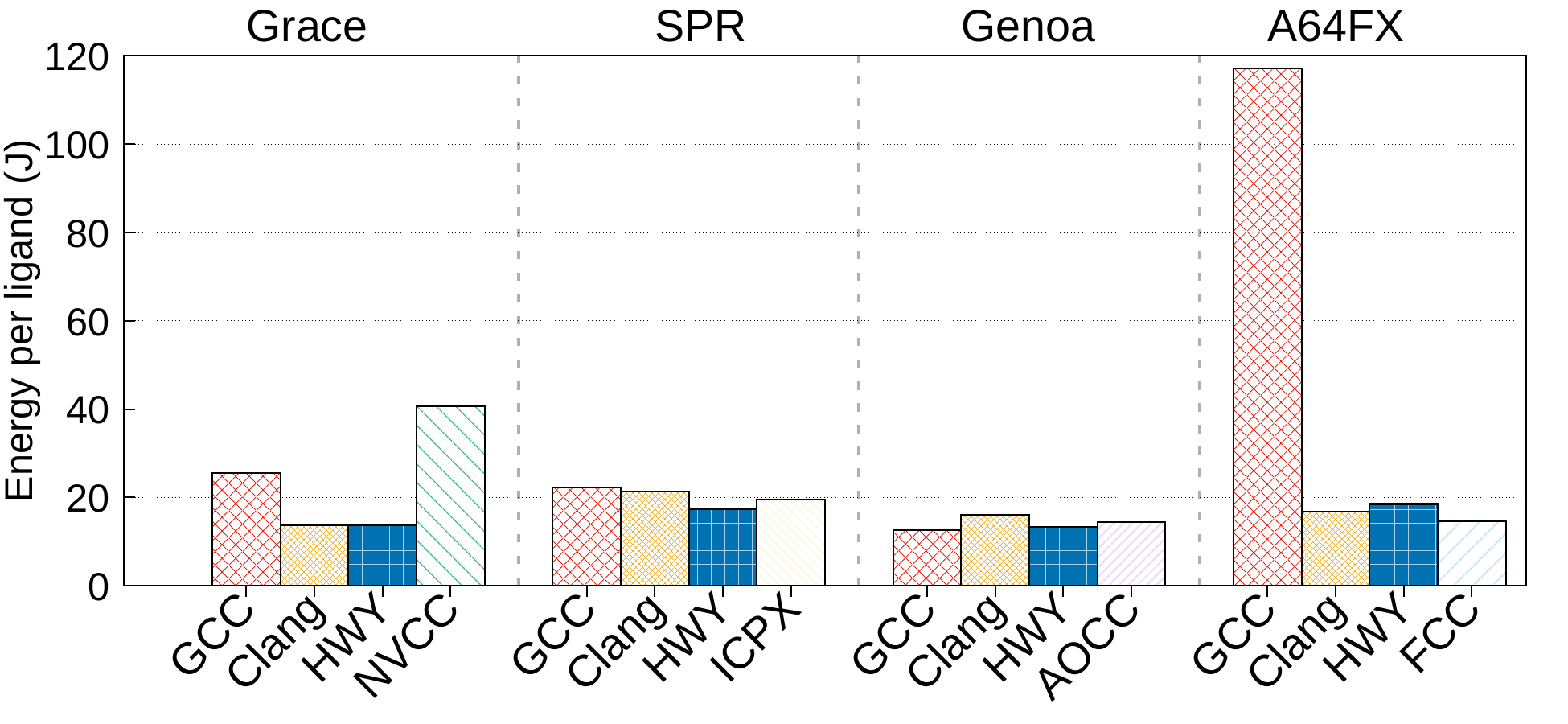}
		\caption{\mudock energy consumption per evaluated ligand (in Joules).}
		\label{figure:energy}
	\end{subfigure}
	\caption{\mudock computational cost analysis on diverse architectures and compilers.}
	\label{figure:tradeoff}
\end{figure*}
Large-scale scientific simulations require significant computational resources in terms of core hours, which can be provided by supercomputing centers.
Therefore, we analyze the tradeoff between the cost and energy consumption per ligand evaluated.
We expect similar energy consumption across architectures, as ARM processors typically consume less power, though their execution times are longer on average compared to x86 processors.
On the other hand, we expect a lower cost per ligand for ARM processors, given their lower energy consumption and operational costs.
Such large-scale campaigns leverage all the computational power offered by computing nodes, and as such, we consider only multi-core scenarios in the following section.
Figure~\ref{figure:energy} shows the energy consumption per ligand in Joules of \mudock.
The reported energy evaluation is conducted on \mudock execution end-to-end, as the energy counters are unreliable and cannot be evaluated using the marker API on a multi-core scenario.
For a similar reason, we could not report Graviton data, as energy counters were inaccessible.
We observed a spike in energy consumption for GCC, as it is not vectorizing effectively, leading to much longer execution times per ligand.
In contrast, the HWY and Clang implementations balance performance and energy consumption across all architectures.
Figure~\ref{figure:cost} shows the cost per ligand evaluated, using the lowest prices offered by cloud providers and HPC centers when writing~\footnote{See references in Table~\ref{tab:cpuFeatures}}.
For example, we considered the lowest price on cloud providers as a standard dedicated instance for 3 years.
As expected, despite comparable energy consumption per ligand, ARM architectures generally offer a lower cost per ligand.
We remark on the differences in cost on the Grace chip since the node hour price also considers the node's GPU.
Notably, the A64FX and Sapphire processor demonstrates the best value for money across all architectures.
Despite the low energy and cost per ligand, ARM architectures overall show a higher execution time, which should be accounted for in the scenario of urgent computing~\cite{uc} when the time to solution matters.

%% file: sections/discussions.tex
\section{Discussion}
\label{sec:discssions}
From our results, adapting HPC applications on modern CPUs requires considering the complex interaction between compilers, architectures, and the target application.
When time-to-solution is critical, the x86 CPUs we tested are faster (higher core count and clock frequency); however, ARM-based CPUs provide a better cost per ligand.
Clang and HWY provide good performance portability, although Clang provides higher performance on ARM than HWY, thanks to ARM's performance libraries, while HWY performs better on x86 architectures.
Despite the technical challenges, we successfully obtained vectorized code on all architectures except for GCC and NVC++ on ARM, due to a GLIBC issue.


Auto-vectorization capabilities of modern compilers allow developers to improve performance across multiple architectures using the same codebase.
However, performance engineers should carefully consider the compilation environment and the compiler they use.
For example, Clang allows users to select the vectorization library that improves vectorization performance on ARM architectures, but Clang does not pick it by default.
GCC relies on the vectorized libraries provided by the system's GLIBC.
Previous studies on ARM architectures~\cite{ookami} highlighted the absence of vectorized math functions in older GLIBC versions.
Recent GLIBC releases have improved support for vectorized functions on ARM, but vectorized GLIBC math functions are not always available on all systems.
As a result, some ARM system installations lack vectorized math libraries, such as \textit{libmvec.so}.
Compilation flags (e.g., \textit{-fopt-info-vec-missed} for GCC and \textit{-Rpass-missed=loop-vectorize} for Clang) help highlight missed vectorization opportunities.
Besides the compilation flag, application development should consider complex control flow analysis and the use of pragmas.
Using pragmas in code helps provide compilers with hints about vectorization requirements.
In particular, Clang offers a set of valuable pragmas to control vectorization, unrolling, and instruction interleaving.
Thus, when optimizing scientific workloads for target architectures, developers should account for available vectorized instructions, out-of-order execution resources, caches, and compilers used.

%% file: sections/conclusions.tex
\section{Conclusion}
\label{sec:conclusions}
Scientific applications usually hinge on HPC centers since the computational power constrains them.
Molecular docking for virtual screening falls in this category, where the number of evaluated ligands depends only on the available time budget .
We focus on the \mudock{} computation kernel that estimates the interaction strength between a ligand and the target protein.
This case study provides valuable insight into common computational patterns of scientific codes, often missing from synthetic benchmarks.
In particular, we compare the auto-vectorization ability of a compiler with respect to explicit vectorization.
Performance portability is a key metric to capture the heterogeneity landscape of HPC centers.
For these reasons, we benchmark seven compilers that target the same code base that uses \textit{\#pragma omp simd} to enable auto-vectorization.
As a reference, we use HWY to emit intrinsics for the target architecture automatically.

We benchmark \mudock{} on diverse HPC CPU architectures.
Experimental results demonstrate that vectorization substantially enhances loop performance across the board, although it remains constrained by architectural resources.
In particular, our results show that x86 CPUs typically offer faster time-to-solution, but ARM and x86 architectures are more similar with respect to cost-to-performance and energy consumption.
However, the interaction between the code, the compilation environment, and architectural features has a complex relationship that can lead to unexpected results.
For example, the lack of a vectorized GLIBC can severely limit the performance on ARM.
For this reason, we provide practical guidelines for developing performance-portable scientific applications in this work.
Even though we evaluate \mudock{}, the guidelines might apply to other scientific applications that share similar computational patterns, specifically, memory lookups into large constant data structures and loop vectorization involving mathematical functions.

%% file: main.bbl
\begin{thebibliography}{10}
\providecommand{\url}[1]{#1}
\csname url@samestyle\endcsname
\providecommand{\newblock}{\relax}
\providecommand{\bibinfo}[2]{#2}
\providecommand{\BIBentrySTDinterwordspacing}{\spaceskip=0pt\relax}
\providecommand{\BIBentryALTinterwordstretchfactor}{4}
\providecommand{\BIBentryALTinterwordspacing}{\spaceskip=\fontdimen2\font plus
\BIBentryALTinterwordstretchfactor\fontdimen3\font minus
  \fontdimen4\font\relax}
\providecommand{\BIBforeignlanguage}[2]{{%
\expandafter\ifx\csname l@#1\endcsname\relax
\typeout{** WARNING: IEEEtran.bst: No hyphenation pattern has been}%
\typeout{** loaded for the language `#1'. Using the pattern for}%
\typeout{** the default language instead.}%
\else
\language=\csname l@#1\endcsname
\fi
#2}}
\providecommand{\BIBdecl}{\relax}
\BIBdecl

\bibitem{bigrun}
\BIBentryALTinterwordspacing
D.~Gadioli, E.~Vitali, F.~Ficarelli, C.~Latini, C.~Manelfi, C.~Talarico,
  C.~Silvano, C.~Cavazzoni, G.~Palermo, and A.~R. Beccari, ``Exscalate: An
  extreme-scale virtual screening platform for drug discovery targeting
  polypharmacology to fight sars-cov-2,'' \emph{IEEE Transactions on Emerging
  Topics in Computing}, vol.~11, no.~1, pp. 170--181, 2023. [Online].
  Available: \url{https://doi.org/10.1109/TETC.2022.3187134}
\BIBentrySTDinterwordspacing

\bibitem{vssummit}
\BIBentryALTinterwordspacing
A.~Acharya, R.~Agarwal, M.~B. Baker, J.~Baudry, D.~Bhowmik, S.~Boehm, K.~G.
  Byler, S.~Y. Chen, L.~Coates, C.~J. Cooper, O.~Demerdash, I.~Daidone, J.~D.
  Eblen, S.~Ellingson, S.~Forli, J.~Glaser, J.~C. Gumbart, J.~Gunnels,
  O.~Hernandez, S.~Irle, D.~W. Kneller, A.~Kovalevsky, J.~Larkin, T.~J.
  Lawrence, S.~LeGrand, S.-H. Liu, J.~Mitchell, G.~Park, J.~Parks, A.~Pavlova,
  L.~Petridis, D.~Poole, L.~Pouchard, A.~Ramanathan, D.~M. Rogers,
  D.~Santos-Martins, A.~Scheinberg, A.~Sedova, Y.~Shen, J.~C. Smith, M.~D.
  Smith, C.~Soto, A.~Tsaris, M.~Thavappiragasam, A.~F. Tillack, J.~V. Vermaas,
  V.~Q. Vuong, J.~Yin, S.~Yoo, M.~Zahran, and L.~Zanetti-Polzi,
  ``Supercomputer-based ensemble docking drug discovery pipeline with
  application to covid-19,'' \emph{Journal of Chemical Information and
  Modeling}, vol.~60, no.~12, pp. 5832--5852, 2020, pMID: 33326239. [Online].
  Available: \url{https://doi.org/10.1021/acs.jcim.0c01010}
\BIBentrySTDinterwordspacing

\bibitem{TOP500}
\BIBentryALTinterwordspacing
{TOP500 Project}, ``{TOP500: The List},'' 2025, accessed: 2025-02-14. [Online].
  Available: \url{https://top500.org/lists/top500/}
\BIBentrySTDinterwordspacing

\bibitem{fugaku}
S.~Matsuoka, ``Fugaku and a64fx: the first exascale supercomputer and its
  innovative arm cpu,'' in \emph{2021 Symposium on VLSI Circuits}, 2021, pp.
  1--3.

\bibitem{lumiC}
\BIBentryALTinterwordspacing
{LUMI Supercomputer}, ``Lumi hardware overview,'' 2023, accessed: 2025-02-17.
  [Online]. Available: \url{https://docs.lumi-supercomputer.eu/hardware/lumic/}
\BIBentrySTDinterwordspacing

\bibitem{ecmwf_supercomputer}
\BIBentryALTinterwordspacing
{European Centre for Medium-Range Weather Forecasts}, ``Supercomputer
  facility,'' 2022, accessed: 2025-04-25. [Online]. Available:
  \url{https://www.ecmwf.int/en/computing/our-facilities/supercomputer-facility}
\BIBentrySTDinterwordspacing

\bibitem{modernCPU}
\BIBentryALTinterwordspacing
S.~Ghose, ``General-purpose multicore architectures,'' 2025. [Online].
  Available: \url{https://arxiv.org/abs/2408.12999}
\BIBentrySTDinterwordspacing

\bibitem{biousage}
\BIBentryALTinterwordspacing
J.~Domke, K.~Matsumura, M.~Wahib, H.~Zhang, K.~Yashima, T.~Tsuchikawa,
  Y.~Tsuji, A.~Podobas, and S.~Matsuoka, ``Double-precision fpus in
  high-performance computing: An embarrassment of riches?'' in \emph{2019 IEEE
  International Parallel and Distributed Processing Symposium (IPDPS)}, 2019,
  pp. 78--88. [Online]. Available:
  \url{https://doi.org/10.1109/IPDPS.2019.00019}
\BIBentrySTDinterwordspacing

\bibitem{Beccari2023}
\BIBentryALTinterwordspacing
A.~Beccari, L.~Dionigi, E.~Nicastri, C.~Manelfi, and E.~Gavioli, \emph{The Drug
  Repurposing Strategy in the Exscalate4CoV Project: Raloxifene Clinical
  Trials}.\hskip 1em plus 0.5em minus 0.4em\relax Cham: Springer International
  Publishing, 2023, pp. 19--26. [Online]. Available:
  \url{https://doi.org/10.1007/978-3-031-30691-4_3}
\BIBentrySTDinterwordspacing

\bibitem{autodock}
\BIBentryALTinterwordspacing
G.~M. Morris, D.~S. Goodsell, R.~S. Halliday, R.~Huey, W.~E. Hart, R.~K. Belew,
  and A.~J. Olson, ``Automated docking using a lamarckian genetic algorithm and
  an empirical binding free energy function,'' \emph{Journal of Computational
  Chemistry}, vol.~19, no.~14, p. 1639–1662, Nov. 1998. [Online]. Available:
  \url{http://dx.doi.org/10.1002/(SICI)1096-987X(19981115)19:14<1639::AID-JCC10>3.0.CO;2-B}
\BIBentrySTDinterwordspacing

\bibitem{googleHighway}
\BIBentryALTinterwordspacing
Google, ``Google highway,'' 2025, accessed: 2025-02-14. [Online]. Available:
  \url{https://github.com/google/highway}
\BIBentrySTDinterwordspacing

\bibitem{gromacs}
\BIBentryALTinterwordspacing
D.~Van Der~Spoel, E.~Lindahl, B.~Hess, G.~Groenhof, A.~E. Mark, and H.~J.~C.
  Berendsen, ``Gromacs: Fast, flexible, and free,'' \emph{Journal of
  Computational Chemistry}, vol.~26, no.~16, pp. 1701--1718, 2005. [Online].
  Available: \url{https://onlinelibrary.wiley.com/doi/abs/10.1002/jcc.20291}
\BIBentrySTDinterwordspacing

\bibitem{openfoam}
\BIBentryALTinterwordspacing
G.~Chen, Q.~Xiong, P.~J. Morris, E.~G. Paterson, A.~Sergeev, and Y.-C. Wang,
  ``Openfoam for computational fluid dynamics,'' \emph{Notices of the American
  Mathematical Society}, vol.~61, no.~4, p. 354, Apr. 2014. [Online].
  Available: \url{http://dx.doi.org/10.1090/noti1095}
\BIBentrySTDinterwordspacing

\bibitem{mudock}
\BIBentryALTinterwordspacing
E.~Polimi, ``mudock,'' 2025, accessed: 2025-02-14. [Online]. Available:
  \url{https://github.com/elvispolimi/muDock}
\BIBentrySTDinterwordspacing

\bibitem{cost}
\BIBentryALTinterwordspacing
S.~Morgan, P.~Grootendorst, J.~Lexchin, C.~Cunningham, and D.~Greyson, ``The
  cost of drug development: A systematic review,'' \emph{Health Policy}, vol.
  100, no.~1, p. 4–17, Apr. 2011. [Online]. Available:
  \url{http://dx.doi.org/10.1016/j.healthpol.2010.12.002}
\BIBentrySTDinterwordspacing

\bibitem{moderndd}
\BIBentryALTinterwordspacing
J.~Hughes, S.~Rees, S.~Kalindjian, and K.~Philpott, ``Principles of early drug
  discovery,'' \emph{British Journal of Pharmacology}, vol. 162, no.~6, p.
  1239–1249, Feb. 2011. [Online]. Available:
  \url{http://dx.doi.org/10.1111/j.1476-5381.2010.01127.x}
\BIBentrySTDinterwordspacing

\bibitem{liu2025impact}
\BIBentryALTinterwordspacing
F.~Liu, O.~Mailhot, I.~S. Glenn, S.~F. Vigneron, V.~Bassim, X.~Xu,
  K.~Fonseca-Valencia, M.~S. Smith, D.~S. Radchenko, J.~S. Fraser, Y.~S. Moroz,
  J.~J. Irwin, and B.~K. Shoichet, ``The impact of library size and scale of
  testing on virtual screening,'' Jul. 2024. [Online]. Available:
  \url{http://dx.doi.org/10.1101/2024.07.08.602536}
\BIBentrySTDinterwordspacing

\bibitem{DDVS}
\BIBentryALTinterwordspacing
M.~Kontoyianni, \emph{Docking and Virtual Screening in Drug Discovery}.\hskip
  1em plus 0.5em minus 0.4em\relax New York, NY: Springer New York, 2017, pp.
  255--266. [Online]. Available:
  \url{https://doi.org/10.1007/978-1-4939-7201-2_18}
\BIBentrySTDinterwordspacing

\bibitem{alphafold}
\BIBentryALTinterwordspacing
J.~Jumper, R.~Evans, A.~Pritzel \emph{et~al.}, ``Highly accurate protein
  structure prediction with alphafold,'' \emph{Nature}, vol. 596, no. 7873, pp.
  583--589, 2021. [Online]. Available:
  \url{https://www.nature.com/articles/s41586-021-03819-2}
\BIBentrySTDinterwordspacing

\bibitem{md}
\BIBentryALTinterwordspacing
X.-Y. Meng, H.-X. Zhang, M.~Mezei, and M.~Cui, ``Molecular docking: A powerful
  approach for structure-based drug discovery,'' \emph{Current Computer
  Aided-Drug Design}, vol.~7, no.~2, p. 146–157, Jun. 2011. [Online].
  Available: \url{http://dx.doi.org/10.2174/157340911795677602}
\BIBentrySTDinterwordspacing

\bibitem{minimdock}
\BIBentryALTinterwordspacing
M.~Thavappiragasam, A.~Scheinberg, W.~Elwasif, O.~Hernandez, and A.~Sedova,
  ``Performance portability of molecular docking miniapp on leadership
  computing platforms,'' in \emph{2020 IEEE/ACM International Workshop on
  Performance, Portability and Productivity in HPC (P3HPC)}.\hskip 1em plus
  0.5em minus 0.4em\relax IEEE, Nov. 2020. [Online]. Available:
  \url{http://dx.doi.org/10.1109/P3HPC51967.2020.00009}
\BIBentrySTDinterwordspacing

\bibitem{autodock4}
\BIBentryALTinterwordspacing
G.~M. Morris, R.~Huey, W.~Lindstrom, M.~F. Sanner, R.~K. Belew, D.~S. Goodsell,
  and A.~J. Olson, ``Autodock4 and autodocktools4: Automated docking with
  selective receptor flexibility,'' \emph{Journal of Computational Chemistry},
  vol.~30, no.~16, pp. 2785--2791, 2009. [Online]. Available:
  \url{https://doi.org/10.1002/jcc.21256}
\BIBentrySTDinterwordspacing

\bibitem{linpack}
\BIBentryALTinterwordspacing
J.~J. Dongarra, P.~Luszczek, and A.~Petitet, ``The linpack benchmark: past,
  present and future,'' \emph{Concurrency and Computation: Practice and
  Experience}, vol.~15, no.~9, pp. 803--820, 2003. [Online]. Available:
  \url{https://doi.org/10.1145/3632261.3632267}
\BIBentrySTDinterwordspacing

\bibitem{openbenchmarking}
\BIBentryALTinterwordspacing
{Phoronix Media}, ``Openbenchmarking.org: Cross-platform, open-source automated
  benchmarking platform,'' 2025, accessed: 2025-04-25. [Online]. Available:
  \url{https://openbenchmarking.org/}
\BIBentrySTDinterwordspacing

\bibitem{benchpark}
\BIBentryALTinterwordspacing
O.~Pearce, A.~Scott, G.~Becker, R.~Haque, N.~Hanford, S.~Brink, D.~Jacobsen,
  H.~Poxon, J.~Domke, and T.~Gamblin, ``Towards collaborative continuous
  benchmarking for hpc,'' in \emph{Proceedings of the SC '23 Workshops of the
  International Conference on High Performance Computing, Network, Storage, and
  Analysis}, ser. SC-W '23.\hskip 1em plus 0.5em minus 0.4em\relax New York,
  NY, USA: Association for Computing Machinery, 2023, p. 627–635. [Online].
  Available: \url{https://doi.org/10.1145/3624062.3624135}
\BIBentrySTDinterwordspacing

\bibitem{speccpu}
\BIBentryALTinterwordspacing
J.~Bucek, K.-D. Lange, and J.~v.~Kistowski, ``Spec cpu2017: Next-generation
  compute benchmark,'' in \emph{Companion of the 2018 ACM/SPEC International
  Conference on Performance Engineering}, ser. ICPE '18.\hskip 1em plus 0.5em
  minus 0.4em\relax New York, NY, USA: Association for Computing Machinery,
  2018, p. 41–42. [Online]. Available:
  \url{https://doi.org/10.1145/3185768.3185771}
\BIBentrySTDinterwordspacing

\bibitem{GenArchBench}
\BIBentryALTinterwordspacing
L.~López-Villellas, R.~Langarita-Benítez, A.~Badouh, V.~Soria-Pardos,
  Q.~Aguado-Puig, G.~López-Paradís, M.~Doblas, J.~Setoain, C.~Kim, M.~Ono,
  A.~Armejach, S.~Marco-Sola, J.~Alastruey-Benedé, P.~Ibáñez, and
  M.~Moretó, ``Genarchbench: A genomics benchmark suite for arm hpc
  processors,'' \emph{Future Generation Computer Systems}, vol. 157, pp.
  313--329, 2024. [Online]. Available:
  \url{https://www.sciencedirect.com/science/article/pii/S0167739X24001250}
\BIBentrySTDinterwordspacing

\bibitem{memoryARM}
\BIBentryALTinterwordspacing
Y.~Kang, S.~Ghosh, M.~Kandemir, and A.~M\'{a}rquez, ``Studying cpu and memory
  utilization of applications on fujitsu a64fx and nvidia grace superchip,'' in
  \emph{Proceedings of the International Symposium on Memory Systems}, ser.
  MEMSYS '24.\hskip 1em plus 0.5em minus 0.4em\relax New York, NY, USA:
  Association for Computing Machinery, 2024, p. 198–207. [Online]. Available:
  \url{https://doi.org/10.1145/3695794.3695813}
\BIBentrySTDinterwordspacing

\bibitem{micrograce}
\BIBentryALTinterwordspacing
J.~Laukemann, G.~Hager, and G.~Wellein, ``Microarchitectural comparison and
  in-core modeling of state-of-the-art cpus: Grace, sapphire rapids, and
  genoa,'' 2024. [Online]. Available: \url{https://arxiv.org/abs/2409.08108}
\BIBentrySTDinterwordspacing

\bibitem{10.1145/3636480.3637097}
\BIBentryALTinterwordspacing
N.~A. Simakov, M.~D. Jones, T.~R. Furlani, E.~Siegmann, and R.~J. Harrison,
  ``First impressions of the nvidia grace cpu superchip and nvidia grace hopper
  superchip for scientific workloads,'' in \emph{Proceedings of the
  International Conference on High Performance Computing in Asia-Pacific Region
  Workshops}, ser. HPCAsia '24 Workshops.\hskip 1em plus 0.5em minus
  0.4em\relax New York, NY, USA: Association for Computing Machinery, 2024, p.
  36–44. [Online]. Available: \url{https://doi.org/10.1145/3636480.3637097}
\BIBentrySTDinterwordspacing

\bibitem{10.1145/3581576.3581621}
\BIBentryALTinterwordspacing
W.~Elwasif, W.~Godoy, N.~Hagerty, J.~A. Harris, O.~Hernandez, B.~Joo, P.~Kent,
  D.~Lebrun-Grandie, E.~Maccarthy, V.~Melesse~Vergara, B.~Messer, R.~Miller,
  S.~Oral, S.~Bastrakov, M.~Bussmann, A.~Debus, K.~Steiniger, J.~Stephan,
  R.~Widera, S.~Bryngelson, H.~Le~Berre, A.~Radhakrishnan, J.~Young,
  S.~Chandrasekaran, F.~Ciorba, O.~Simsek, K.~Clark, F.~Spiga, J.~Hammond,
  S.~John, D.~Hardy, S.~Keller, J.-G. Piccinali, and C.~Trott, ``Application
  experiences on a gpu-accelerated arm-based hpc testbed,'' in
  \emph{Proceedings of the HPC Asia 2023 Workshops}, ser. HPCAsia '23
  Workshops.\hskip 1em plus 0.5em minus 0.4em\relax New York, NY, USA:
  Association for Computing Machinery, 2023, p. 35–49. [Online]. Available:
  \url{https://doi.org/10.1145/3581576.3581621}
\BIBentrySTDinterwordspacing

\bibitem{SYCLARM}
\BIBentryALTinterwordspacing
H.~Liang, C.~Deng, P.~Zhang, J.~Fang, T.~Tang, and C.~Huang, ``An empirical
  performance evaluation of sycl on arm multi-core processors,'' \emph{CCF
  Transactions on High Performance Computing}, Feb. 2025. [Online]. Available:
  \url{http://dx.doi.org/10.1007/s42514-024-00212-z}
\BIBentrySTDinterwordspacing

\bibitem{kokkos}
\BIBentryALTinterwordspacing
D.~Sahasrabudhe, E.~T. Phipps, S.~Rajamanickam, and M.~Berzins, ``A portable
  simd primitive using kokkos for heterogeneous architectures,'' in
  \emph{Accelerator Programming Using Directives}, S.~Wienke and
  S.~Bhalachandra, Eds.\hskip 1em plus 0.5em minus 0.4em\relax Cham: Springer
  International Publishing, 2020, pp. 140--163. [Online]. Available:
  \url{https://doi.org/10.1007/978-3-030-49943-3_7}
\BIBentrySTDinterwordspacing

\bibitem{poenaru2021evaluation}
\BIBentryALTinterwordspacing
A.~Poenaru, T.~Deakin, S.~McIntosh-Smith, S.~Hammond, and A.~Younge,
  ``\BIBforeignlanguage{English}{An evaluation of the fujitsu a64fx for hpc
  applications},'' in \emph{\BIBforeignlanguage{English}{Cray User Group
  2021}}, May 2021, cray User Group 2021 ; Conference date: 03-05-2021 Through
  05-05-2021. [Online]. Available: \url{https://cug.org/cug-2021/}
\BIBentrySTDinterwordspacing

\bibitem{ficarelliAsia}
\BIBentryALTinterwordspacing
F.~Barbari, F.~Ficarelli, and D.~Cesarini, ``High-throughput drug discovery on
  the fujitsu a64fx architecture,'' in \emph{Proceedings of the International
  Conference on High Performance Computing in Asia-Pacific Region Workshops},
  ser. HPCAsia '24 Workshops.\hskip 1em plus 0.5em minus 0.4em\relax New York,
  NY, USA: Association for Computing Machinery, 2024, p. 17–23. [Online].
  Available: \url{https://doi.org/10.1145/3636480.3637095}
\BIBentrySTDinterwordspacing

\bibitem{ljavx}
\BIBentryALTinterwordspacing
H.~Watanabe and K.~M. Nakagawa, ``Simd vectorization for the lennard-jones
  potential with avx2 and avx-512 instructions,'' \emph{Computer Physics
  Communications}, vol. 237, pp. 1--7, 2019. [Online]. Available:
  \url{https://www.sciencedirect.com/science/article/pii/S001046551830376X}
\BIBentrySTDinterwordspacing

\bibitem{poenaru2021performance}
\BIBentryALTinterwordspacing
A.~Poenaru, W.-C. Lin, and S.~McIntosh-Smith, ``A performance analysis of
  modern parallel programming models using a compute-bound application,'' in
  \emph{International Conference on High Performance Computing}.\hskip 1em plus
  0.5em minus 0.4em\relax Springer, 2021, pp. 332--350. [Online]. Available:
  \url{https://doi.org/10.1007/978-3-030-78713-4_18}
\BIBentrySTDinterwordspacing

\bibitem{avx}
\BIBentryALTinterwordspacing
C.~Lomont, ``Introduction to intel advanced vector extensions,'' \emph{Intel
  white paper}, vol.~23, pp. 1--21, 2011. [Online]. Available:
  \url{https://hpc.llnl.gov/sites/default/files/intelAVXintro.pdf}
\BIBentrySTDinterwordspacing

\bibitem{sve}
\BIBentryALTinterwordspacing
N.~Stephens, S.~Biles, M.~Boettcher, J.~Eapen, M.~Eyole, G.~Gabrielli,
  M.~Horsnell, G.~Magklis, A.~Martinez, N.~Premillieu, A.~Reid, A.~Rico, and
  P.~Walker, ``The arm scalable vector extension,'' \emph{IEEE Micro}, vol.~37,
  no.~2, pp. 26--39, 2017. [Online]. Available:
  \url{https://doi.org/10.1109/MM.2017.35}
\BIBentrySTDinterwordspacing

\bibitem{ookami}
\BIBentryALTinterwordspacing
M.~A.~S. Bari, B.~Chapman, A.~Curtis, R.~J. Harrison, E.~Siegmann, N.~A.
  Simakov, and M.~D. Jones, ``A64fx performance: experience on ookami,'' in
  \emph{2021 IEEE International Conference on Cluster Computing (CLUSTER)},
  2021, pp. 711--718. [Online]. Available:
  \url{https://doi.org/10.1109/Cluster48925.2021.00106}
\BIBentrySTDinterwordspacing

\bibitem{autovectorization}
\BIBentryALTinterwordspacing
J.~G. Feng, Y.~P. He, and Q.~M. Tao, ``Evaluation of compilers’ capability of
  automatic vectorization based on source code analysis,'' \emph{Scientific
  Programming}, vol. 2021, no.~1, p. 3264624, 2021. [Online]. Available:
  \url{https://doi.org/10.1155/2021/3264624}
\BIBentrySTDinterwordspacing

\bibitem{stdsimd}
\BIBentryALTinterwordspacing
I.~JTC1/SC22/WG21, ``Working draft, standard for programming language c++,''
  July 2019, iSO/IEC JTC1/SC22/WG21 Working Draft. [Online]. Available:
  \url{https://wg21.link/N4808}
\BIBentrySTDinterwordspacing

\bibitem{xsimd}
xtensor-stack developers, ``xsimd: C++ wrappers for simd intrinsics,''
  \url{https://github.com/xtensor-stack/xsimd}, 2016, accessed: 2025-04-22.

\bibitem{pp}
\BIBentryALTinterwordspacing
S.~Pennycook, J.~Sewall, and V.~Lee, ``Implications of a metric for performance
  portability,'' \emph{Future Generation Computer Systems}, vol.~92, pp.
  947--958, 2019. [Online]. Available:
  \url{https://www.sciencedirect.com/science/article/pii/S0167739X17300559}
\BIBentrySTDinterwordspacing

\bibitem{openmp}
\BIBentryALTinterwordspacing
O.~A.~R. Board, \emph{OpenMP Application Programming Interface Specification
  5.2}, 2021. [Online]. Available: \url{https://www.openmp.org/specifications/}
\BIBentrySTDinterwordspacing

\bibitem{trott2010autodock}
\BIBentryALTinterwordspacing
O.~Trott and A.~J. Olson, ``Autodock vina: improving the speed and accuracy of
  docking with a new scoring function, efficient optimization, and
  multithreading,'' \emph{Journal of computational chemistry}, vol.~31, no.~2,
  pp. 455--461, 2010. [Online]. Available:
  \url{https://doi.org/10.1002/jcc.21334}
\BIBentrySTDinterwordspacing

\bibitem{autodockGPU}
\BIBentryALTinterwordspacing
D.~Santos-Martins, L.~Solis-Vasquez, A.~F. Tillack, M.~F. Sanner, A.~Koch, and
  S.~Forli, ``Accelerating autodock4 with gpus and gradient-based local
  search,'' \emph{Journal of Chemical Theory and Computation}, vol.~17, no.~2,
  pp. 1060--1073, 2021, pMID: 33403848. [Online]. Available:
  \url{https://doi.org/10.1021/acs.jctc.0c01006}
\BIBentrySTDinterwordspacing

\bibitem{autodockVINAGPU}
\BIBentryALTinterwordspacing
J.~Ding, S.~Tang, Z.~Mei, L.~Wang, Q.~Huang, H.~Hu, M.~Ling, and J.~Wu,
  ``Vina-gpu 2.0: Further accelerating autodock vina and its derivatives with
  graphics processing units,'' \emph{Journal of Chemical Information and
  Modeling}, vol.~63, no.~7, pp. 1982--1998, 2023, pMID: 36941232. [Online].
  Available: \url{https://doi.org/10.1021/acs.jcim.2c01504}
\BIBentrySTDinterwordspacing

\bibitem{autodockFPGA}
\BIBentryALTinterwordspacing
L.~Solis-Vasquez and A.~Koch, ``A case study in using opencl on fpgas: Creating
  an open-source accelerator of the autodock molecular docking software,'' in
  \emph{FSP Workshop 2018; Fifth International Workshop on FPGAs for Software
  Programmers}, 2018, pp. 1--10. [Online]. Available:
  \url{https://ieeexplore.ieee.org/document/8470463}
\BIBentrySTDinterwordspacing

\bibitem{approx}
\BIBentryALTinterwordspacing
Y.~Zhang, G.~Accordi, D.~Gadioli, and G.~Palermo, ``Harnessing
  quality-throughput trade-off in scoring functions for extreme-scale virtual
  screening campaigns,'' \emph{Future Generation Computer Systems}, vol. 172,
  p. 107863, 2025. [Online]. Available:
  \url{https://www.sciencedirect.com/science/article/pii/S0167739X2500158X}
\BIBentrySTDinterwordspacing

\bibitem{autogrid}
\BIBentryALTinterwordspacing
M.~Olšák, J.~Filipovič, and M.~Prokop, ``Fastgrid -- the accelerated
  autogrid potential maps generation for molecular docking,'' \emph{Computing
  and Informatics}, vol.~29, no.~6+, p. 1325–1336, Jan. 2012. [Online].
  Available: \url{https://www.cai.sk/ojs/index.php/cai/article/view/146}
\BIBentrySTDinterwordspacing

\bibitem{openFOAMlt}
\BIBentryALTinterwordspacing
M.~Votsmeier, ``Efficient implementation of detailed surface chemistry into
  reactor models using mapped rate data,'' \emph{Chemical Engineering Science},
  vol.~64, no.~7, pp. 1384--1389, 2009. [Online]. Available:
  \url{https://www.sciencedirect.com/science/article/pii/S0009250908006532}
\BIBentrySTDinterwordspacing

\bibitem{lammps}
\BIBentryALTinterwordspacing
A.~P. Thompson, H.~M. Aktulga, R.~Berger, D.~S. Bolintineanu, W.~M. Brown,
  P.~S. Crozier, P.~J. {in 't Veld}, A.~Kohlmeyer, S.~G. Moore, T.~D. Nguyen,
  R.~Shan, M.~J. Stevens, J.~Tranchida, C.~Trott, and S.~J. Plimpton, ``Lammps
  - a flexible simulation tool for particle-based materials modeling at the
  atomic, meso, and continuum scales,'' \emph{Computer Physics Communications},
  vol. 271, p. 108171, 2022. [Online]. Available:
  \url{https://www.sciencedirect.com/science/article/pii/S0010465521002836}
\BIBentrySTDinterwordspacing

\bibitem{wrf}
\BIBentryALTinterwordspacing
{WRF Community}, ``Weather research and forecasting (wrf) model,'' 2000.
  [Online]. Available: \url{http://www2.mmm.ucar.edu/wrf/users/}
\BIBentrySTDinterwordspacing

\bibitem{AWSPrice}
AWS, ``Aws calculator,'' \url{ https://calculator.aws }, 2025, accessed:
  2025-02-12.

\bibitem{sprSpec}
\BIBentryALTinterwordspacing
{Intel Corporation}, \emph{Intel® Xeon® Scalable Processors (Formerly
  Sapphire Rapids)}, 2025, accessed: 2025-02-13. [Online]. Available:
  \url{https://www.intel.com/content/www/us/en/ark/products/codename/126212/products-formerly-sapphire-rapids.html}
\BIBentrySTDinterwordspacing

\bibitem{GenoaSpec}
H.~Mujtaba, ``Amd epyc genoa zen 4 cpu lineup specs and benchmarks leaked,''
  https://wccftech.com/amd-epyc-genoa-cpu-lineup-specs-benchmarks-leak-up-to-2-6x-faster-than-intel-xeon/?view=html,
  2022.

\bibitem{neoverse}
C.~Lam, ``Arm’s neoverse v2, in aws’s graviton 4,'' \url{
  https://chipsandcheese.com/p/arms-neoverse-v2-in-awss-graviton-4 }, 2024,
  accessed: 2025-02-12.

\bibitem{FugakuPrice}
Riken, ``Fugaku usage fee,'' \url{
  https://www.hpci-office.jp/en/using_hpci/proposal_submission_current/fugaku_price
  }, 2019, accessed: 2025-02-12.

\bibitem{fujitsuA64FX}
Fujitsu, ``A64fx,'' \url{https://github.com/fujitsu/A64FX/tree/master}, 2025,
  accessed: 2025-02-12.

\bibitem{alps}
CSCS, ``Cscs nvidia grace usage cost,''
  \url{https://2go.cscs.ch/offering/other_customers/}, 2024, accessed:
  2025-02-12.

\bibitem{GraceGuide}
\BIBentryALTinterwordspacing
{NVIDIA Corporation}, \emph{NVIDIA Grace Performance Tuning Guide}, 2025,
  accessed: 2025-02-13. [Online]. Available:
  \url{https://docs.nvidia.com/grace-perf-tuning-guide/system.html}
\BIBentrySTDinterwordspacing

\bibitem{sprRegisters}
\BIBentryALTinterwordspacing
{Chips and Cheese}, ``Golden cove’s vector register file: Checking with
  official spr data,'' 2024, accessed: 2025-02-12. [Online]. Available:
  \url{https://chipsandcheese.com/p/golden-coves-vector-register-file-checking-with-official-spr-data}
\BIBentrySTDinterwordspacing

\bibitem{spr}
C.~Lam, ``Popping the hood on golden cove,'' \url{
  https://chipsandcheese.com/p/popping-the-hood-on-golden-cove}, 2021,
  accessed: 2025-02-12.

\bibitem{zen4}
C.~L. Frontend and E.~Enginem, ``Amd’s zen 4 part 1: Frontend and execution
  engine,'' \url{
  https://chipsandcheese.com/p/amds-zen-4-part-1-frontend-and-execution-engine
  }, 2022, accessed: 2025-02-12.

\bibitem{likwid}
\BIBentryALTinterwordspacing
J.~Treibig, G.~Hager, and G.~Wellein, ``Likwid: A lightweight
  performance-oriented tool suite for x86 multicore environments,'' in
  \emph{2010 39th International Conference on Parallel Processing Workshops},
  2010, pp. 207--216. [Online]. Available:
  \url{https://doi.org/10.1109/ICPPW.2010.38}
\BIBentrySTDinterwordspacing

\bibitem{numa}
\BIBentryALTinterwordspacing
L.~Zaourar, M.~Benazouz, A.~Mouhagir, C.~Falquez, A.~Portero, N.~Ho, E.~Suarez,
  P.~Petrakis, M.~Marazakis, F.~Sgherzi, I.~Fernandez, R.~Dolbeau, and
  D.~Pleiter, ``Case studies on the impact and challenges of heterogeneous numa
  architectures for hpc,'' in \emph{Architecture of Computing Systems}, D.~Fey,
  B.~Stabernack, S.~Lankes, M.~Pacher, and T.~Pionteck, Eds.\hskip 1em plus
  0.5em minus 0.4em\relax Cham: Springer Nature Switzerland, 2024, pp.
  251--265. [Online]. Available:
  \url{https://doi.org/10.1007/978-3-031-66146-4_17}
\BIBentrySTDinterwordspacing

\bibitem{nvidiaPT}
\BIBentryALTinterwordspacing
{NVIDIA}, ``Nvidia grace power and thermals,'' 2023, accessed: 2025-02-17.
  [Online]. Available:
  \url{https://docs.nvidia.com/grace-perf-tuning-guide/power-thermals.html}
\BIBentrySTDinterwordspacing

\bibitem{mediate}
\BIBentryALTinterwordspacing
G.~Vistoli, C.~Manelfi, C.~Talarico, A.~Fava, A.~Warshel, I.~V. Tetko,
  R.~Apostolov, Y.~Ye, C.~Latini, F.~Ficarelli, G.~Palermo, D.~Gadioli,
  E.~Vitali, G.~Varriale, V.~Pisapia, M.~Scaturro, S.~Coletti, D.~Gregori,
  D.~Gruffat, E.~Leija, S.~Hessenauer, A.~Delbianco, M.~Allegretti, and A.~R.
  Beccari, ``Mediate - molecular docking at home: Turning collaborative
  simulations into therapeutic solutions,'' \emph{Expert Opinion on Drug
  Discovery}, vol.~18, no.~8, pp. 821--833, 2023, pMID: 37424369. [Online].
  Available: \url{https://doi.org/10.1080/17460441.2023.2221025}
\BIBentrySTDinterwordspacing

\bibitem{pdbbind}
\BIBentryALTinterwordspacing
R.~Wang, X.~Fang, Y.~Lu, C.-Y. Yang, and S.~Wang, ``The pdbbind database:
  Methodologies and updates,'' \emph{Journal of Medicinal Chemistry}, vol.~48,
  no.~12, pp. 4111--4119, 2005, pMID: 15943484. [Online]. Available:
  \url{https://doi.org/10.1021/jm048957q}
\BIBentrySTDinterwordspacing

\bibitem{fscaling}
{LLVM Project}, ``[avx512] prefering 512-bit vectors on recent intel cpus,''
  \url{https://github.com/llvm/llvm-project/issues/102047}, 2023, gitHub issue
  \#102047, accessed 2025-05-07.

\bibitem{sse}
\BIBentryALTinterwordspacing
H.~Jeong, S.~Kim, W.~Lee, and S.-H. Myung, ``Performance of sse and avx
  instruction sets,'' 2012. [Online]. Available:
  \url{https://arxiv.org/abs/1211.0820}
\BIBentrySTDinterwordspacing

\bibitem{loopfission}
\BIBentryALTinterwordspacing
T.~Odajima, Y.~Kodama, M.~Tsuji, M.~Matsuda, Y.~Maruyama, and M.~Sato,
  ``Preliminary performance evaluation of the fujitsu a64fx using hpc
  applications,'' in \emph{2020 IEEE International Conference on Cluster
  Computing (CLUSTER)}, 2020, pp. 523--530. [Online]. Available:
  \url{https://doi.org/10.1109/CLUSTER49012.2020.00075}
\BIBentrySTDinterwordspacing

\bibitem{a64fxSWP}
\BIBentryALTinterwordspacing
M.~Arai, N.~Fukumoto, and H.~Murai, ``Introducing software pipelining for the
  a64fx processor into llvm,'' in \emph{Proceedings of the International
  Conference on High Performance Computing in Asia-Pacific Region Workshops},
  ser. HPCAsia '24 Workshops.\hskip 1em plus 0.5em minus 0.4em\relax New York,
  NY, USA: Association for Computing Machinery, 2024, p. 1–6. [Online].
  Available: \url{https://doi.org/10.1145/3636480.3637093}
\BIBentrySTDinterwordspacing

\bibitem{scagrace}
\BIBentryALTinterwordspacing
Y.~Kang, S.~Ghosh, M.~Kandemir, and A.~M\'{a}rquez, ``Studying cpu and memory
  utilization of applications on fujitsu a64fx and nvidia grace superchip,'' in
  \emph{Proceedings of the International Symposium on Memory Systems}, ser.
  MEMSYS '24.\hskip 1em plus 0.5em minus 0.4em\relax New York, NY, USA:
  Association for Computing Machinery, 2024, p. 198–207. [Online]. Available:
  \url{https://doi.org/10.1145/3695794.3695813}
\BIBentrySTDinterwordspacing

\bibitem{pprevised}
\BIBentryALTinterwordspacing
S.~J. Pennycook and J.~D. Sewall, ``Revisiting a metric for performance
  portability,'' in \emph{2021 International Workshop on Performance,
  Portability and Productivity in HPC (P3HPC)}, 2021, pp. 1--9. [Online].
  Available: \url{https://doi.org/10.1109/P3HPC54578.2021.00004}
\BIBentrySTDinterwordspacing

\bibitem{uc}
\BIBentryALTinterwordspacing
D.~Gadioli, G.~Accordi, J.~Krenek, M.~Golasowski, L.~Foltyn, J.~Martinovic,
  A.~R. Beccari, and G.~Palermo, ``A portable drug discovery platform for
  urgent computing,'' \emph{Procedia Computer Science}, vol. 240, pp. 42--51,
  2024, proceedings of the First EuroHPC user day. [Online]. Available:
  \url{https://www.sciencedirect.com/science/article/pii/S1877050924016958}
\BIBentrySTDinterwordspacing

\end{thebibliography}
